\documentclass[pdflatex,sn-aps]{sn-jnl}

\usepackage{graphicx}%
\usepackage{multirow}%
\usepackage{amsmath,amssymb,amsfonts}%
\usepackage{amsthm}%
\usepackage{mathrsfs}%
\usepackage[title]{appendix}%
\usepackage{textcomp}%
\usepackage{manyfoot}%
\usepackage{booktabs}%
\usepackage{algorithm}%
\usepackage{algorithmicx}%
\usepackage{algpseudocode}%
\usepackage{listings}%
\usepackage{hyperref}
\usepackage[nameinlink,poorman]{cleveref}
\usepackage[table,xcdraw]{xcolor}
\usepackage{placeins}
\usepackage{color}
\usepackage{colortbl}
\usepackage{pdfpages}
\usepackage{verbatim}

\theoremstyle{thmstyleone}%
\theoremstyle{thmstyletwo}%
\theoremstyle{thmstylethree}%
\crefname{equation}{Eq.}{Eqs.}
\crefname{figure}{Fig.}{Figs.}
\crefname{table}{Table}{Tables} 
\crefname{section}{Section}{Sections}
\crefname{chapter}{Chapter}{Chapters}
\crefname{appendix}{Appendix}{Appendices}
\crefname{algorithm}{Algorithm}{Algorithms}
\crefname{theorem}{Theorem}{Theorems}
\crefname{defn}{Definiton}{Definitions}
\crefname{azm}{Assumption}{Assumptions}
\crefname{corollary}{Corollary}{Corollaries}
\crefname{lemma}{Lemma}{Lemmas}
\crefname{thmprop}{property}{properties}
\crefname{proposition}{Proposition}{Propositions}
\crefname{remark}{Remark}{Remarks}

\def \srtabs {4915} 
\def \srftr {169} 
\def \srbad {123} 
\def \srgood {46} 
\def \srsyn {16} 

\definecolor{b0}{cmyk}{0.90,0.44,0.16,0.18}
\definecolor{b1}{cmyk}{0.22,0.04,0.02,0.00}
\definecolor{b2}{cmyk}{0.42,0.06,0.02,0.00}
\definecolor{b4}{cmyk}{0.61,0.12,0.01,0.00}
\definecolor{b3}{cmyk}{0.91,0.49,0.23,0.35}
\definecolor{b5}{cmyk}{0.90,0.58,0.31,0.53}

\usepackage{carlito}
\usepackage[T1]{fontenc}

\begin{document}

\title{Quantum Machine Learning for Digital Health? A Systematic Review}

\author*[1]{\fnm{Riddhi S.} \sur{Gupta}}\email{riddhi.gupta@uq.edu.au}
\author[1]{\fnm{Carolyn E.} \sur{Wood}}
\author[2]{\fnm{Teyl} \sur{Engstrom}}
\author[2]{\fnm{Jason} D. \sur{Pole}}
\author[1,2]{\fnm{Sally} \sur{Shrapnel}}

\affil*[1]{\orgdiv{School of Mathematics and Physics}, \orgname{The University of Queensland}, \orgaddress{ \city{St Lucia}, \postcode{4067}, \state{QLD}, \country{Australia}}}

\affil[2]{\orgdiv{QDHeC. Centre for Health Services Research. Faculty of Medicine}, \orgname{The University of Queensland}, \orgaddress{ \city{Herston}, \postcode{4006}, \state{QLD}, \country{Australia}}}

\abstract{With the digitization of health data, the growth of electronic health and medical records lowers barriers for using algorithmic techniques for data analysis. While classical machine learning (ML) techniques for health data approach commercialization and incorporation into clinical workflows, there is not yet clear evidence whether quantum machine learning (QML) will actually provide any empirical advantage for digital health data processing tasks. In this systematic literature review we assess whether, in light of developing digital health technologies, quantum machine learning algorithms have the potential to outperform existing classical methods in efficacy or efficiency. We include digital  electronic health/medical records (EH/MRs) and data considered to be a reasonable proxy to EH/MRs. Eligible QML algorithms must be designed for quantum computing hardware, as opposed to quantum-inspired techniques. PubMed, Embase, IEEE, Scopus and a physics preprint server arXiv yielded \srtabs{} studies between 2015 to 10 June 2024. After screening \srftr{} eligible studies, most studies contained widespread technical misconceptions about QML and we excluded \srbad{} studies for insufficient rigor in analysis. Of the remaining \srgood{} studies, only \srsyn{} studies consider realistic QML operating conditions, either by testing algorithms on quantum hardware, or using noisy quantum circuits when assessing QML algorithms. We find QML applications in digital health focus primarily on clinical decision support rather than health service delivery or public health. Nearly all QML models are linear quantum models, and therefore represent a subset of general quantum algorithm structures. Meanwhile, novel data-encoding strategies do not address scalability issues, except in regimes requiring restrictive assumptions about quantum hardware, rendering these protocols inefficient for the general encoding of large health datasets. Our work establishes the current state of evidence for QML-based health applications, paving the way for meaningful QML use-case discovery in digital health.
}

\keywords{quantum machine learning, quantum computing algorithms, digital health, electronic health records, electronic medical records, medical imaging, machine learning, routinely collected health data}

\maketitle
\clearpage
\newpage

\section{Introduction}\label{sec:intro}

Recent years have seen a proliferation of research proclaiming the utility of quantum machine learning (QML) algorithms for analyzing classical data in many sectors, e.g. finance, cybersecurity, logistics, pharmaceuticals, energy, minerals, and healthcare. With the increasing digitization of health data, the growth of electronic health and medical records  \cite{Hecht2019Sep} paves the way for the use of algorithmic techniques – quantum or classical – for analyzing this data. Potential digital health applications could include clinical decision support, clinical predictive health and health monitoring, public health applications and improving health services delivery and data fusion \cite{StanfordWhitePaperEHR2018, Mohsen2022Oct,Obermeyer2016Sep,Rajkomar2018May}. The potential for use-case discovery for QML in healthcare \cite{Ullah2024Jan} and biomedical \cite{Maheshwari2022Jul} applications is found to be compelling in previous systematic reviews. Other broader reviews on quantum computing for health, biology and lifesciences \cite{Flother2023Jan,Cordier2022Nov,Emani2021Jul,Outeiral2021Jan,Basu2023Jul, Biamonte2017Sep,Marchetti2022Nov,Baiardi2023Jul} hypothesize the potential utility of QML algorithms or quantum subroutines in health, but none of these works are rigorous systematic reviews (and thus reproducible). Indeed, across all of these standard and systematic reviews, we find that the strength of the current evidence base even under mildly realistic operating conditions is not examined.

Characterizing the role of QML algorithms applied to real-world classical data is nuanced and a challenging question in applications development but also in fundamental QML theory \cite{Bowles2024Mar,Bermejo2024Aug}. Quantum advantage refers to asymptotic reduction in computational resources (or some other metric \cite{Schuld2022Jul}) required by quantum algorithms when compared to classical counterparts, i.e. resources are saved as problem size scales to infinity. Empirical quantum advantage \cite{Krunic2022} colloquially refers to finite-sized simulations or experiments using quantum over classical algorithms to perform a task, where one assumes any desired resource-savings will scale to larger problems, e.g. in qubit number, high-dimensional or highly structured datasets. However, for classical datasets of arbitrary structure such as those encountered in healthcare settings, there is no known theoretically provable quantum advantage \cite{Schuld2022Jul}. Instead, the field relies on mostly empirical analysis of QML performance for a variety of pseudo-real-world data, where performance differentials between quantum and classical methods on these smaller problems constitute evidence for testing empirical quantum advantage. Most computational analysis of scaling behavior assumes ideal operating conditions and it is unknown if QML methods will retain any benefits in realistic operating settings, such as on near-term noisy quantum hardware. In some cases, the role of quantum algorithms for solving inference tasks has been entirely replaced by equivalent classical capability, in a process known as dequantization (e.g. \cite{Schreiber2023Sep,Sweke2023Sep}). 

In this work, we undertake a systematic literature review of QML applications in digital health between 2015 and 2024. As typical in medical research settings, a systematic literature review is a standard methodological approach for assessing the strength of evidence for proposed interventions in clinical contexts and public health \cite{SRinHealth2024Jun}. Based on existing evidence in literature, we ask whether quantum machine learning algorithms potentially outperform existing classical methods in efficacy or efficiency for digital health. Our objective is to assess the strength of the evidence and dominant trends associated with using QML algorithms for digital health, including assessing the extent to which performance robustness of proposed QML algorithms has been characterized. 

Our current-state analysis reveals that the empirical evidence for QML in digital health cannot conclusively address our research question. We find that numerous studies had to be excluded due to a lack of technical rigor in their analysis of QML algorithms. The majority of eligible studies use only ideal simulations of QML algorithms, thereby excluding the resource overhead incurred for error-mitigated or error-corrected algorithms required for noisy quantum hardware. Of high quality studies, nearly all QML algorithms are found to linear quantum models, and therefore represent a small subset of general QML.  Most use-cases in digital health focussed on providing clinical support, and no studies considered health service delivery or public health applications. Only two synthesized studies used electronic health records for quantum machine learning applications, while the remaining studies repeatedly gravitated towards a handful of open-source health databases. Finally, 13 studies used quantum hardware demonstrations and separated into two classes: either algorithms for a gate-based, universal quantum computer using up to 20 qubits, or quantum annealers using $O(100)$ qubits. Whether potential advantages of QML can be retained in the presence of noise is largely unaddressed in all studies.

The structure of this document is as follows. We begin by providing an overview of quantum machine learning in \cref{sec:bkgd}. In \cref{sec:methods}, we outline our approach in accordance with systematic review procedures and guidelines documenting database selection, screening procedures and inclusion criteria for articles. We tabulate key research themes and conduct meta-analysis of our final set of articles in \cref{sec:results}. Finally, we return to address the research questions of our review, and comment on research opportunities, gaps, limitations and future work in \cref{sec:discn} before providing a future outlook and concluding remarks in \cref{sec:fo} and \cref{sec:con} respectively.

\section{Overview of quantum machine learning (QML)}\label{sec:bkgd}

Quantum computing refers to a broad category of algorithms, for which it is desired that quantum computing hardware will be required to perform some of the computations. Quantum machine learning algorithms are a subset of quantum computation. For the scope of this review, the input of a quantum algorithm is associated with a classical dataset, and an inference problem is defined on the classical dataset. 

Quantum computational advantage accrues when a quantum algorithm can reduce the number of operations required to solve this inference problem as the size of the problem becomes asymptotically large. Here, the problem size is typically associated with features of the input data e.g. with input data dimension. From a computer science perspective, algorithms can either improve on the number of queries or samples required (sample complexity) or the number of parallelizable quantum operations (time complexity, or runtime). When quantum algorithms enable improvements in either sample or time complexity, this is sometimes informally referred to as 'quantum advantage' or ‘speed-up'. An additional metric of memory complexity quantifies the size or type of classical data structures required to store classical inputs, outputs and efficiently recall intermediary classical information during computation, but memory complexity is typically not encountered in the literature for quantum algorithms. A comparison of computational costs of selected classical vs. quantum algorithms for ideal mathematical regimes can be found in Ref.~\cite{Outeiral2021Jan}, but these were not encountered for real-world health data in our review.

In this section, we provide background to common families of QML algorithms. Quantum algorithms separate into two different categories in this review: gate-based quantum models, or quantum annealing. This categorization can broadly reflect the difference between digital and universal vs. analogue and non-universal quantum computing. Background quantum notation and a fuller discussion is provided in \cref{sec:app:qnotation}. We conclude this section by outlining the role of classical data in quantum machine learning.

\subsection{Algorithms for universal, gate-based quantum computers}

A large subset of quantum algorithms are designed for gate-based universal quantum computers. These algorithms include quantum kernel methods (including quantum support vector machines), quantum neural networks, quantum convolutional neural networks, and quantum deep learning. We summarize these quantum models by considering how outputs are generated from inputs.

\begin{figure}[htbp]
\centering
\includegraphics[width=0.95\textwidth]{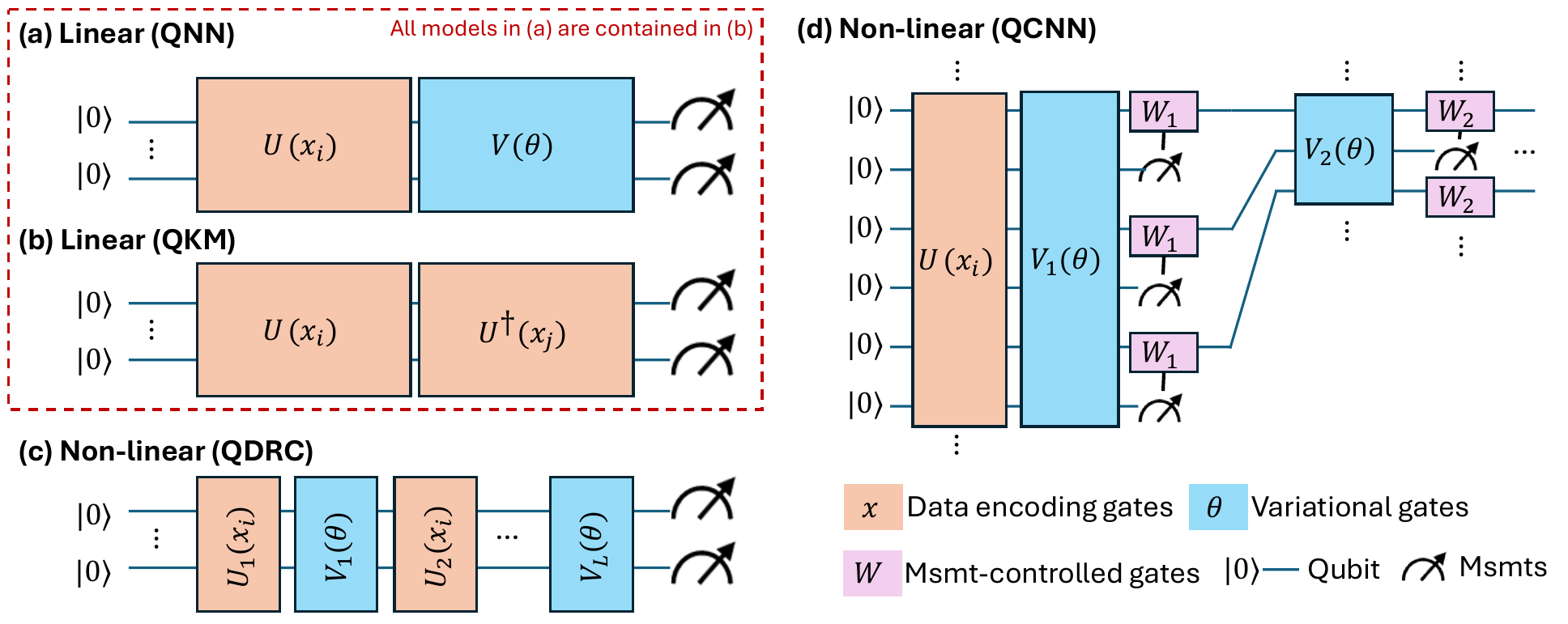}
\caption{Quantum circuit depictions of linear vs. non-linear embedding of classical data in quantum models \cite{Jerbi2023Jan}. Horizontal wires represent qubits where input states are shown as ket $|\cdot \rangle$ symbols; temporal order of computations progresses from left to right. Boxed quantum gates (blue, orange) are reversible rotations, or `unitary gates', of quantum states, i.e. $U^\dagger = U^{-1}$. If data encoding (blue) is separable from variational gates (orange), then the quantum model is linear. `Circuit size' refers to the number of qubits, while `circuit depth' represents the number of time-steps required to run the full circuit assuming that quantum operations on disjoint qubits have been parallelized. (a)-(c) Measurements (msmts.) are pushed to the end; quantum circuit can be summarized by a unitary operation. (d) Mid-circuit measurement outcomes change quantum operations `on the fly' (pink). Circuits with tunable $\theta$ (blue gates) can be broadly referred to as variational (VQC) or parameterized (PQC) quantum circuits.} \label{fig:qc}
\end{figure}

QML algorithms are frequently represented as quantum circuits, with some examples in \cref{fig:qc}. In this visualization, a quantum algorithm is composed of input qubit states denoted with ket-notation $|\cdot\rangle$ and boxed operations denote quantum gates. These gates are associated with reversible, logical operations performed on quantum states. The circuit is terminated with measurements of a quantum state which yield probabilistic outcomes, `0' or `1', where probabilities are determined by the quantum circuit. Suppose for some input quantum state,  $\rho_0$, the average output of a quantum algorithm is given by $f(x, \theta)$ where $(x, \theta)$ define classical inputs to a quantum algorithm. Here, $\rho_0$ represents an input state, such as all qubits in their ground (zero) state; $x$ represents one sample of real data with dimension $d$, $x \in \mathbb{R}^d$ for a dataset with $N$ samples; and tunable free parameters, $\theta$, that parameterize the circuit. One encodes data $x$ into quantum states using a parameterized quantum gate, denoted $U(x)$. The choice of a data encoder, $U(x)$, is discussed in \cref{sec:roleofdata}. Meanwhile, free parameters, $\theta$, implement classically optimized or trained quantum gates, $V(\theta)$. With these assumptions, the desired output information required from the algorithm is typically given with reference to an observable quantity, $\hat{O}$. This output information is inherently statistical, i.e. one must infer average information about $\hat{O}$ from a statistical ensemble of `0' or `1' measurements obtained by repeatedly preparing and measuring the same quantum circuit $N_{s}$ number of times. Therefore to extract information about $\hat{O}$, we build up an ensemble of quantum measurements by repeatedly running a quantum circuit  $N_{s}$ number of times for a single instance of $x$, and repeating for different choices of $x$. 

A quantum machine learning algorithm typically consists of input data ($x$-dependent) and tunable ($\theta$-dependent) quantum operations. Using \cref{sec:app:qnotation}, we can write the general output of a QML algorithm as,
\begin{align}
    f(x, \theta) &:=  \mathrm{Tr} \left[ U(x, \theta) \rho_0 U^\dagger(x, \theta)  \hat{O} \right] = \langle \rho_{x,\theta}, \hat{O} \rangle, 
\end{align} where the data ($x$-dependent) and tunable ($\theta$-dependent) components of the quantum state $\rho_{x,\theta}$ cannot be separated. In the above, $U(x, \theta)$ represents a parameterized quantum gate which depends on data $x$ and tunable parameters $\theta$.  The output of a QML algorithm thus computes the overlap between information in the quantum state $ \rho_{x,\theta} = U(x, \theta) \rho_0 U^\dagger(x, \theta)$, and the desired output $\hat{O}$, using an inner product. 

In contrast, linear quantum models allow us to separate the $x$-dependent quantum operations and $\theta$-dependent quantum operations within the inner product \cite{Jerbi2023Jan}. In these models, we perform data encoding operations followed by tunable gates $V(\theta)$. As shown in \cref{fig:qc}(a), a linear quantum neural network (QNN) can be expressed by, 
\begin{align}
    f(x, \theta) &:=  \mathrm{Tr} \left[ V(\theta) U(x) \rho_0 U^\dagger(x) V^\dagger(\theta) \hat{O} \right] = \langle \rho_x, \hat{O}_\theta \rangle.
\end{align} In the above, $\theta$ can take the form of any other classical parameters that are not $x$; data encoding is expressed by $\rho_x := U(x) \rho_0 U^\dagger(x)$, and the parameterized neural net is expressed as $\hat{O}_\theta:= V^\dagger(\theta) \hat{O} V(\theta)$. We note that the embedding $U(x)$ can be nonlinear transformation of the input data, $x$. However, the terminology `linear' quantum model refers to the linearity of the model with respect to the embedding i.e. data-dependent and parameterized components of the quantum algorithm can be separated as shown above.

With this structure, we can additionally describe many other types of quantum machine learning algorithms. For example, we can omit $\theta$ entirely, and recover sophisticated algorithms that focus on data encoding procedures. In quantum kernel methods (QKMs), $\theta$ is replaced by training data, and the algorithm output $f$ during prediction represents a linear combination of all training samples. Sometimes the action of $\rho$, $U(x)$ or $V(\theta)$ is non-trivially restricted to some subset of quantum states. Quantum convolutional neural networks (QCNNs), quantum generative adversarial networks, quantum causal modelling, quantum transformers, and quantum deep reinforcement learning all have regimes in which they reduce to linear quantum models as discussed in \cref{sec:app:qnotation}. 

\subsection{Algorithms for adiabatic quantum computers}

Quantum annealing algorithms assume a very specific type of quantum computing hardware, namely adiabatic computers,  (e.g. D-Wave) to solve specific learning tasks. Adiabatic quantum computers can approximately solve computationally hard (i.e. `NP-hard') problems \cite{Date2021May} including approximately solving combinatorial optimization problems. The main class of problems encountered in this review relates to quadratic unconstrained binary optimization (QUBO). Examples of QUBO optimization problems include  regression, classification, and data compression tasks. Classical, quantum and hybrid annealers can all approximately solve QUBO optimization problems \cite{Guddanti2023}, or be used to draw samples from particular types of probability distributions (e.g. Boltzmann distributions) \cite{Piat2018}. While more general forms of adiabatic quantum computing than annealing techniques do exist, we did not encounter any within our included literature, and for this reason have not included a discussion of this form of learning.

Quantum algorithms for QUBO formulations have provable advantage over classical counterparts in some regimes. Quantum QUBO algorithms for optimizing support vector machines (SVMs) and balanced k-means clustering have better computational complexities compared to classical counterparts; while quantum algorithms for QUBO formulations of regression have equivalent computational complexity to classical algorithms \cite{Date2021May}. For this limited class of problems, quantum adiabatic computers, such as D-Wave 2X processors, can access $\approx$ 1000 qubits, which is an order of magnitude larger than $O(100)$ qubit processors for universal non-annealing quantum computers developed by IBM and Google.

\bigskip

The subsections above provide a high-level summary of classes of quantum algorithms that were encountered in this review, but cannot be construed as a comprehensive overview of quantum machine learning (see for example, Refs. \cite{Wang2024Jan}, \cite{Peral-Garcia2024Feb}). We also note that it is possible to realize quantum annealing tasks on gate-based quantum computers, e.g. Ref. \cite{McKeever2024Jun}, and therefore our classification represents one choice of a non-exclusive method for framing the discussion of QML algorithms. In the next subsection, we consider methods for classical data input to QML algorithms. 

\subsection{Data encoding and preprocessing}\label{sec:roleofdata}

So far we have introduced quantum machine learning algorithms in generality without reference to the dataset under consideration.  In this section, we study inputs for QML and  distinguish between data encoding vs. data-preprocessing steps. Data encoding describes the process of representing classical data as quantum states. Data encoding is required for both annealing and non-annealing quantum algorithms.  On the other hand, data pre-processing is concerned with using classical techniques to clean up, rescale, compress, or transform data prior to quantum data encoding, or using a classical algorithm. Both data encoding and pre-processing are discussed below.

Characteristics of classical data and the representation of this data in a quantum algorithm can affect potential attainability of computational advantage in solving inference tasks \cite{Schuld2021Mar,Sweke2021Mar}. Ideally, data encoders must be efficient in computational complexity in both circuit size (number of qubits) and circuit depth (number of parallel operations). There are a number of ways to embed classical data $x$ in quantum states, as summarized in \cref{fig:tab:encoding}. For continuous variable inputs, one may use binary representation of data to finite precision $\tau$ and encode using discrete methods such as basis encoding, as reported in \cref{fig:tab:encoding}. The growth of the number of computations required for encoding is mathematically expressed in $\mathcal{O}(g(n))$-notation to express an upper bound $g(n)$ on the number of operations as the argument $n$ goes to infinity, ignoring constant multiplicative or additive factors.  As an example from \cref{fig:tab:encoding}, angle-encoding can be prepared in constant depth but scales linearly with number of qubits. The trade-off is switched for amplitude encoding, which in general scale linearly with runtime and logarithmically with qubit number. 

\begin{figure}[htbp]
\centering
{\small \carlitoOsF \begin{tabular} {  m{3cm} m{3cm} m{3cm} m{3cm}} 
\rowcolor{b0}\color{white} Encoding & \color{white} Circuit Size (\# of Qubits) & \color{white} Circuit Depth (Runtime)  & \color{white} Ref. \\
\hline
Angle encoding &  $O(d)$ & $O(1)$ & \cite{Weigold2021Dec}\\
\hline
Basis encoding &  $O(d\tau)$ & $O(1),O(Nd\tau)$ & \cite{Weigold2021Dec,Schuld2018SQML} \\
\hline
Amplitude encoding &  $O(\log_2(d\tau))$ &  $O(\log_2(Nd\tau)),O(Nd\tau)$ &\cite{Weigold2021Dec,Schuld2018SQML} \\
\hline
QRAM  & $ O(d\tau + m) $ & $O(\log(m))$ & $m = O(d\tau)$ \cite{Giovannetti2008Apr,Chen2023Mar}\\
\hline
Parallel unary loader & $O(\sqrt{d})$ & $O(\sqrt{d} \log_2 (d))$ & \cite{Johri2021Aug} \\
\bottomrule
\end{tabular}}
\caption{Table of data encoding strategies with circuit size and depth scaling for a data vector $x \in \mathbb{R}^d$ with dimensionality $d$, a total number of $N$ samples in the database. Suppose the binary representation of a single element of $x$ is of length $\tau$ bits, with an additional bit for storing signs $\pm 1$. Then the binary representation $b$ of $x$ has length $d(\tau + 1) = O(d\tau)$. Only amplitude encoding has sub-polynomial scaling of number of qubits with number of classical bits required to represent input data. Runtime complexity refers to the number of parallel quantum operations during encoding. For basis encoding, a single data vector can be encoded in constant time \cite{Leymann2020Sep} but linear time is required to prepare a superposition over $N$ samples \cite{Ventura1998Jul}. For amplitude encoding, runtime is linear for general datasets but can be reduced to sub-polynomial scaling only for restricted datasets or by enabling an algorithm to access QRAM using an additional $m$ ancillary qubits under idealized conditions \cite{Schuld2018SQML,Chen2023Mar}.
\label{fig:tab:encoding}}
\end{figure}

Hardware-specific considerations can change exact implementation details of a quantum algorithm, but generally do not change computational resource requirements reported in \cref{fig:tab:encoding}. The decomposition of required quantum operations to the native set of quantum gates available on hardware may change the number of operations, e.g. replacing one 2-qubit gate with a decomposition involving several single and 2-qubit gates. Similarly, hardware implementation of any continuous variable often also incurs finite precision. In most cases, these changes are multiplicative  or additive with problem size. These multiplicative or additive changes do not affect the overall asymptotic scaling behaviour of the encoder. Some data encoders are not intended as a near-term, implementable strategy. For example, quantum random access memories (QRAM) \cite{Giovannetti2008Apr} use an additional $m = O(d\tau)$ ancillary qubits to randomly access superpositions of basis-encoded states in favourable logarithmic $O(\log(m))$ time. However, robust QRAMs remain extremely challenging to implement on hardware \cite{Phalak2023May}. Finally, the parallel unary encoder assumes specific hardware capabilities that affect the complexity of data encoding, as discussed in \cref{sec:discn}.

Since data encoding is expensive in quantum resources, and may impact performance,  raw data is often pre-processed before encoding. This pre-processing can have many goals, e.g. to compress raw data, identify key features, or address missing values.  For most near-term demonstrations of QML, it is well known that dimensionality reduction of classical datasets is often required to encode data into small or intermediate scale quantum circuits. 

\section{Methods}\label{sec:methods}

Having discussed broad categories of QML algorithms, we now turn to the research question that determines the design of our systematic review. Unlike most systematic reviews, the use of QML algorithms in health settings encompasses a broad range of clinical applications. We use the SPICE framework \cite{Booth2006Jul} to ask: 
\begin{center}
    \fbox{ \parbox{\textwidth}{In developing digital health technologies, could quantum machine learning algorithms potentially outperform existing classical methods in efficacy or efficiency?}}
\end{center} A systematic review was conducted in line with the PRISMA (Preferred Reporting Items for Systematic Reviews and Meta-Analyses) \cite{PRISMA2024Aug} (\cref{sec:app:primsa}) and was registered on PROSPERO (ID: CRD42024562024) \cite{PROSPERO}. Screening and data extraction were performed in Covidence \cite{Covidence}. Commonly used nomenclature encountered in this review is summarized in \cref{sec:app:nomenclature}.
 
\begin{figure}[htbp]
\centering
\includegraphics[clip, trim=1.3cm 4.5cm 1.3cm 2.1cm, width=1\textwidth]{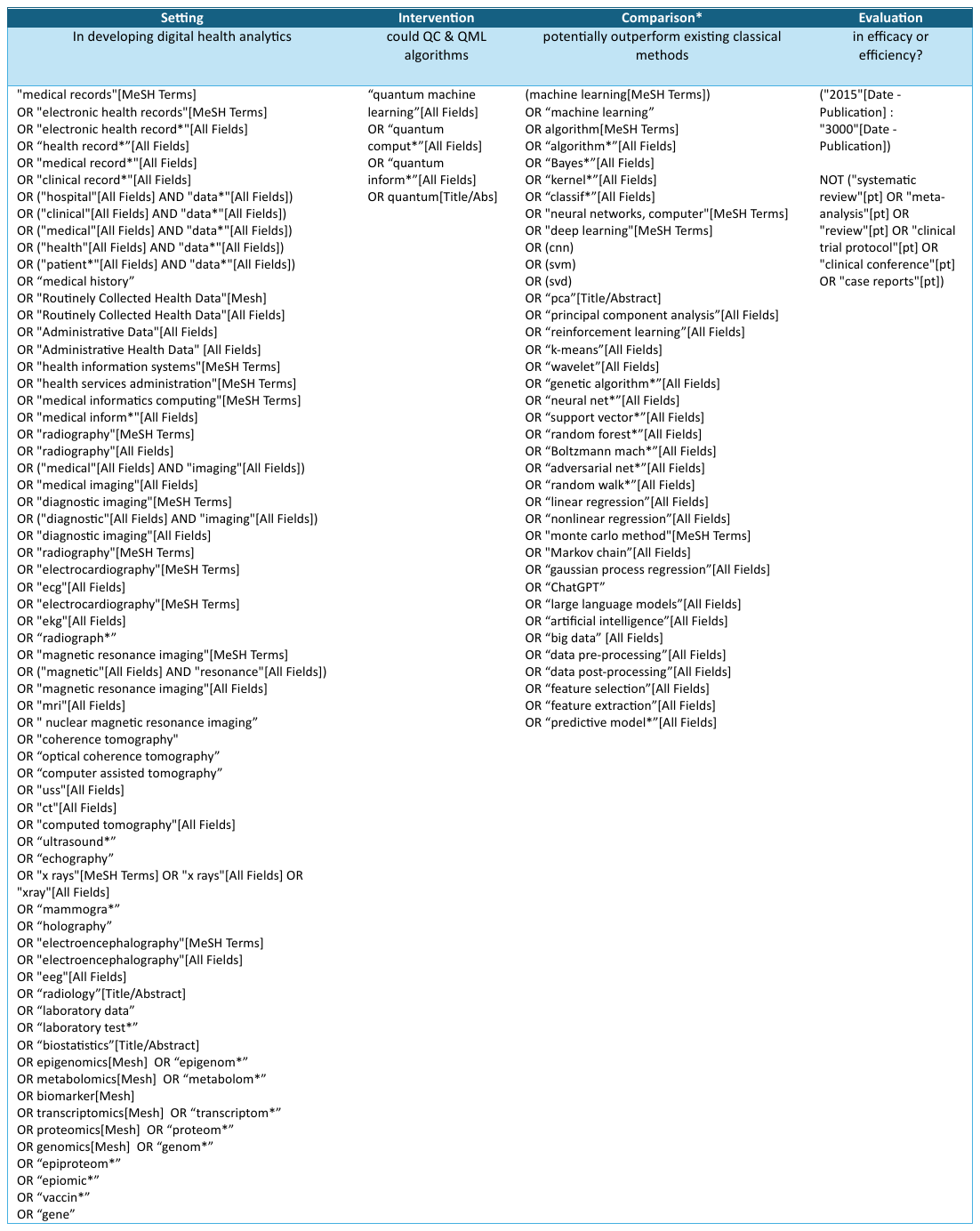}
\caption{Search strategy in the SPICE framework \cite{Booth2006Jul}. Columns represent digital health setting and perspective (S,P), quantum intervention (I), classical comparator (C), and study characteristics for evidence evaluation (E). Search concepts within columns are combined with logical `OR' statements, while independent columns are concatenated with logical `AND' statements. Search concept performance differs by database e.g. PubMed, Embase (shown above) support reliable MESH explosion of health concepts; while Scopus, arXiv and IEEE instead rely on root word truncation. Complete search syntax modifications by database are enclosed in \cref{sec:app:search}. }\label{fig:tab:spice} 
\end{figure}

{\bf Search strategy:} Our search strategy is formed by decomposing our research question into elements of the SPICE framework \cite{Booth2006Jul}, as summarized in \cref{fig:tab:spice}. Only articles published after 2015 were included, as the first commercially-available quantum computer was made accessible in 2016 \cite{IBMQExperience} and digitization of health information into electronic records \cite{Hecht2019Sep} is relatively recent. Hence both factors prohibit meaningful applications development prior to this date. Search syntax was refined by trial and error on PubMed (\cref{fig:tab:spice}) in consultation with a health research librarian, and adapted to other databases (Embase, Scopus, arXiv and IEEE, refer \cref{sec:app:search}). Key articles were identified as litmus tests to sense check database-specific search term strategies. Searches were conducted from 10 May to 10 June 2024. 

\begin{figure}[htbp]
\centering
\includegraphics[clip, trim=1.3cm 6.3cm 1.3cm 2.2cm, width=1\textwidth]{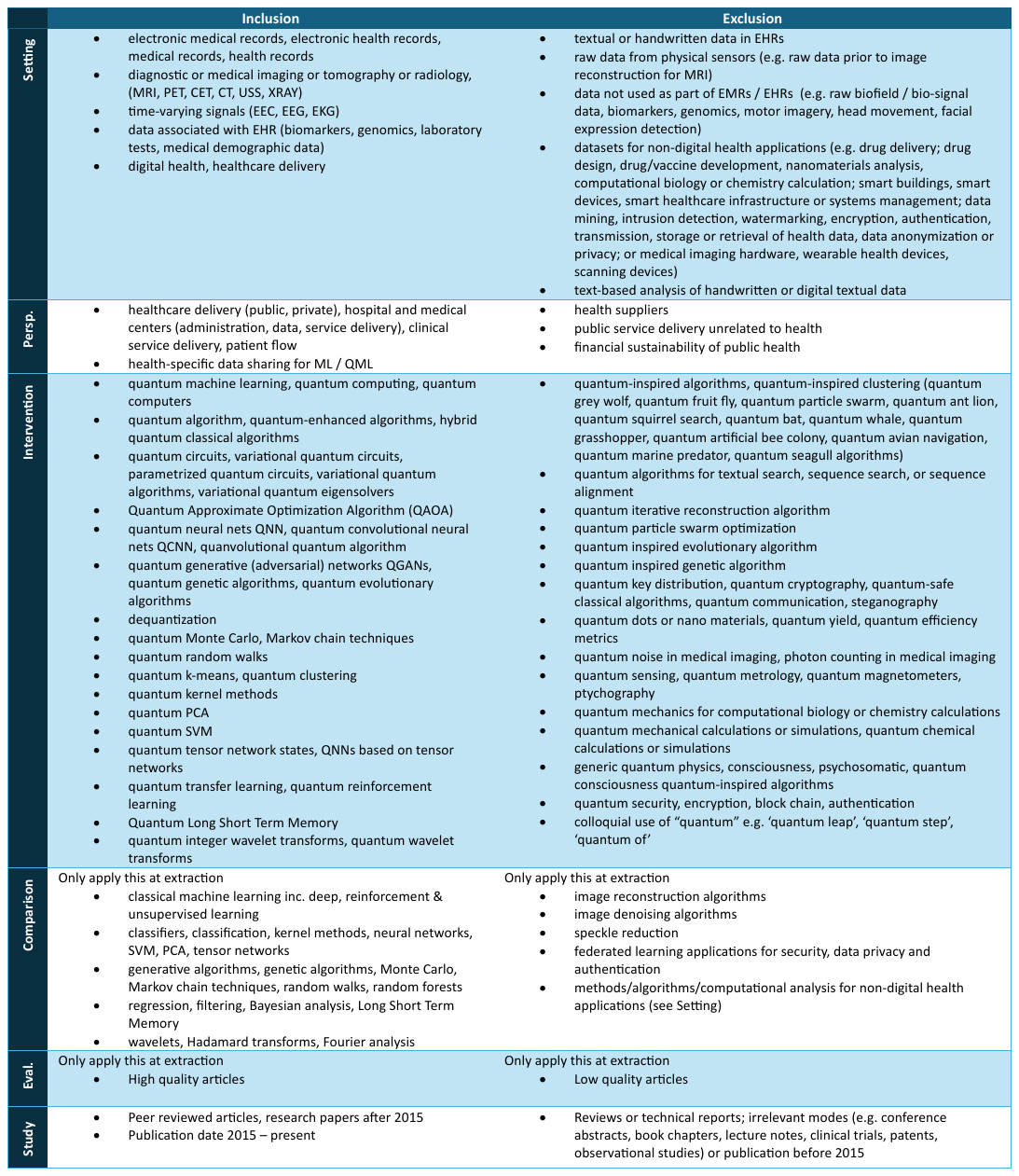}
\caption{Eligibility criteria for title and abstract screening and full text review.  Columns represent inclusion/exclusion criteria, while rows are elements of SPICE framework aligned with search strategy in \cref{fig:tab:spice}. Criteria for digital health setting, quantum interventions, and study characteristics were applied during title and abstract screen and full text review. Meanwhile, criteria for  classical comparators and evaluating technical robustness were only applied after data extraction and prior to synthesis.}\label{fig:tab:eligibility} 
\end{figure}

{\bf Inclusion/exclusion criteria:} The eligibility criteria for the screening process is summarized in \cref{fig:tab:eligibility}. Our study setting prioritized digital health data sources that consist of electronic medical records (EMRs) or electronic health records (EHRs). EMRs represent a real-time patient health record that collects, stores, and displays clinical information as the foundation of a digital hospital as opposed to an EHR which displays summarized patient information to the consumer in the community and across multiple health care providers. The terms “EMR” and “EHR” may be used interchangeably in some countries.  Since health data is subject to strict privacy and security legislation, we also consider data that could be reasonably considered to be in an EHR or EMR, thereby permitting the inclusion of open source and published health datasets that are typically used for proof-of-principle results in both classical and quantum ML. While EHR/EMR data typically includes medical imaging, laboratory data, time-varying signals and patient information, we also include genomics data and biomarkers when used in a context where they supplement a patient’s EHR or EMR for diagnosis or predictive health applications. A notable exclusion is textual search or analysis of digital or handwritten clinical notes, as these would imply looking at an entirely different class of algorithms that have little or no overlap with unstructured data analysis of non-textual health datasets listed above.

Our criteria also prioritized QML algorithms that were genuinely intended to be run on quantum computing hardware, and at least aspired to demonstrate some kind of advantageous scaling property as the number of qubits is increased. In \cref{fig:tab:eligibility}, we list the sheer number of algorithms that are classical computations with a nominal usage of the word `quantum’.  This list was added to throughout screening as new terms were encountered. Many studies technically obfuscated the distinction between QML algorithms and classical computations that use quantum mechanical theory or other insights. For instance, in medical imaging we exclude quantum mechanical corrections to classical algorithms which help to reduce noise in reconstructing images from raw sensor data. Finally, we exclude quantum algorithms unlikely to arise in the context of analyzing classical digital health data, for example: quantum sensing, quantum cryptography, and quantum algorithms for genome pattern matching, genomic sequence alignment, or molecular and chemistry simulations.

{\bf Screening:} Two independent researchers conducted title and abstract screening of all search results: one reviewer had a health background, while the other had a physics background. Full text review was performed by a total of three reviewers. For consistency, one reviewer participated for all screening stages including both abstract and full text screening. Conflicts were resolved through internal discussion or by involving a third reviewer’s opinion.

{\bf Data extraction and study quality appraisal:}  Study characteristics were extracted for all included studies. Additionally, a study quality appraisal was performed to form consensus-based decisions about including or excluding particular studies based on robustness \cite{Popay2006Jan}. These appraisals are typically implemented during data extraction and prior to narrative synthesis \cite{Popay2006Jan}. Our study quality assessment criteria analyses the rigor with which QML algorithms were investigated \cite{Bowles2024Mar} and we do not include a myriad of other potential benchmarks, e.g. for clinical robustness. At least two reviewers independently scored eligible studies, and the maximum score over both reviewers was selected during consensus formation. Attributes of low vs. high quality studies with were compared with respect to our criteria. Full data extraction template is enclosed in \cref{sec:app:dbextract} and extracted data as well as underlying analysis code for data extraction is available online \cite{Gupta2024QML}.

\section{Results}\label{sec:results}

In this section, we present results data from our systematic review in two stages. Firstly, we depict results of the screening process and the study quality appraisal, which has led to a focus on \srsyn{} studies for final synthesis. Secondly, we summarize synthesized evidence and discuss the extent to which our original research question is addressed.

\begin{figure}[htbp]
\centering
\includegraphics[width=0.6\textwidth]{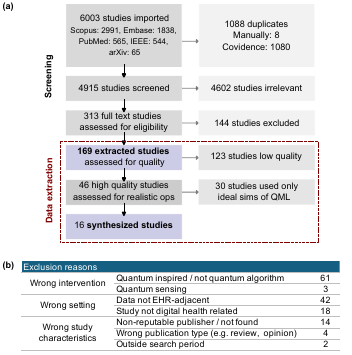}
\caption{PRISMA diagram. (a) Overview of systematic review from screening to synthesis. Of \srtabs{} studies returned from search in 5  databases, 93.6\% of abstracts are deemed ineligible based on digital health setting and quantum intervention criteria. During full text review, 144 studies are excluded, and exclusion reasons are given in (b). Data is extracted (red box) for \srftr{}  eligible studies; additional quality assessment and selection criteria is applied to yield a total of \srsyn{} studies for final synthesis. \label{fig:prisma}} 
\end{figure}

\begin{figure}[htbp]
\centering
{\small \carlitoOsF \begin{tabular} {  m{45em}  cm{1cm} } 
\rowcolor{b0}\color{white} Study Quality Appraisal & \color{white} Counts \\
\hline
\rowcolor{b1} Q1: Explains quantum algorithm selection by referencing learning problem class or dataset structure &  \\
\hline
0 : No theory or empirical rationale for quantum algorithm selection discussed or cited & 32 \\
1 : Quantum algorithm selection is empirical or mostly cites empirical literature & 115 \\
2 : Quantum algorithm selection is linked to underlying class of learning problem or data structure & 21 \\
3 : Quantum algorithm has provable advantage with respect to class of learning problem or data structure & 1 \\
\hline
\rowcolor{b1} Q2: Identifies/discusses impact of data encoding on quantum algorithm performance &  \\
\hline
0 : Encoding methodology omitted or incomplete & 38 \\
0 : Impact of different encoding strategies on overall performance is not analyzed & 85 \\
1 : Performance impact is discussed with incomplete analysis (e.g. compares at least 2 methods) & 37 \\
2 : Performance impact is well characterized empirically or theoretically & 7 \\
Not applicable  & 2 \\
\hline
\rowcolor{b1} Q3: Identifies/discusses impact of classical input data processing on quantum algorithm performance &  \\
\hline
0 : Data pre-processing methodology is omitted or incomplete & 18 \\
0 : Impact of different data pre-processing strategies on overall performance is not analyzed & 96 \\
1 : Performance impact is discussed with incomplete analysis (e.g. compares at least 2 methods) & 45 \\
2 : Performance impact is well characterized empirically or theoretically & 7 \\
Not applicable (no classical input data processing) & 3 \\
\hline
\rowcolor{b1} Q4: (EMPIRICAL ONLY) Dimensionality of data input for quantum algorithm &  \\
\hline
0 : Not reported or discussed; or unclear & 31 \\
0 : Negligible i.e. O(1) & 102 \\
1 : Small i.e. O(10) & 30 \\
2 : Intermediate i.e. $O(10^2)$ & 3 \\
3 : Large i.e. $O(10^3)$ or greater & 2 \\
Not applicable (theory study) & 1 \\
\bottomrule
\end{tabular}}
\caption{ Resulting score distribution for study quality assessment applied to \srftr{} eligible studies at extraction. Each study was scored independently by two reviewers. Consensus scores reflect agreed values formed by discussion with a view to taking the maximum possible of two scores where differences in interpretation were trivial. \label{fig:tab:qascore}}
\end{figure}

\begin{figure}[htbp]
\centering
\includegraphics[width=1.0\textwidth]{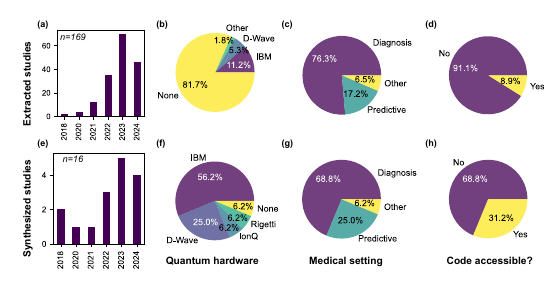}
\caption {Data extraction for eligible studies in \cref{fig:prisma}, with $n=169, 16$ for number of  studies extracted ((a)-(d)) or synthesized ((e)-(h)) respectively.  Synthesized studies exclude eligible studies that do not meet our quality criteria or rely solely on ideal simulations. (a), (e) Histogram of number of studies by publication year; eligible studies show rapid increase in number by year in (a) while synthesized studies appear more uniformly distributed in (e). (b),(f) Percentage of studies that mentioned using quantum hardware using either health or non-health datasets. `None' in (b) indicates usage of both ideal and noisy simulations, while `None' in (f) indicates only noisy simulations of quantum computers. (c),(g) Percentage of studies summarized by digital health setting, broadly characterized as aiding clinical diagnosis (`Diagnosis'), enabling predictive health (`Predictive'), and `Other'. (d), (h) Number of studies providing open access to software code and input datasets. The absence of code/data availability statements; statements `upon reasonable request'; broken links or unusable repositories are all categorized as `No' above.  \label{fig:tab:metadata}}
\end{figure}

\subsection{Characterization of synthesized studies}

Our systematic review is summarized by a PRISMA diagram in  \cref{fig:prisma}(a). Our searches identified \srtabs{} distinct studies. A total of 313 studies passed title and abstract screening and went through full-text screening, of which \srftr{} met eligibility criteria. Inter-rater reliability as measured by Cohen's kappa was substantial for title and abstract screening (0.72) and moderate for full text screening (0.48). According to the distribution of exclusion reasons shown in \cref{fig:prisma}(b), the most frequent cause for full-text exclusion during screening was distinguishing between genuinely quantum algorithms designed to run on quantum hardware, vs. classical computation invoking ideas, insights or jargon from quantum physics. If the health setting of physics-centric QML studies is not explicit (e.g. for instance, \cite{Benedetti2021Oct}), then these studies will not be returned in search nor pass title and abstract screening for any systematic review.

To address issues of technical rigor, we approach data extraction in two steps: first, the application of the quality assessment criteria, and secondly, narrowing the focus to studies that investigated realistic operating conditions either via noisy simulations or by testing algorithms on real quantum hardware.  The distribution of quality scores after consensus is reported in \cref{fig:tab:qascore}. Only 6 of \srftr{} studies led to non-trivial and unresolvable differences in scoring criteria between two independent reviewers, indicating 96.4\% consensus rate for quality assessment scoring. Borderline studies arise when a quality score for two of the following three concerns remains unresolved by reviewers: insufficient performance analysis of classical pre-processing before data input to quantum algorithm, insufficient performance analysis of scalability using qubit numbers $> O(1)$, and/or insufficient performance analysis of choice of data encoding strategy. 

The resulting metadata for extracted (synthesized) studies is shown in the top (bottom) row of \cref{fig:tab:metadata}. Of all eligible studies in \cref{fig:tab:metadata} (top row), 138 of 169 (81.7\%)  use only simulations of quantum machine learning applications for digital health without testing on hardware. Where simulations are the only evidence base in a study, only 7 out of 138 studies use some form of noisy simulations, while the remaining 131 studies use only ideal simulations. When restricting to synthesized studies in \cref{fig:tab:metadata} (bottom row),  a greater proportion of studies do appear to test quantum algorithms on actual quantum hardware (refer \cref{fig:tab:metadata} (b) vs. (f)).

\begin{figure}[htbp]
\centering
\includegraphics[clip, trim=1.75cm 7.1cm 1.75cm 6cm, width=1.\textwidth]{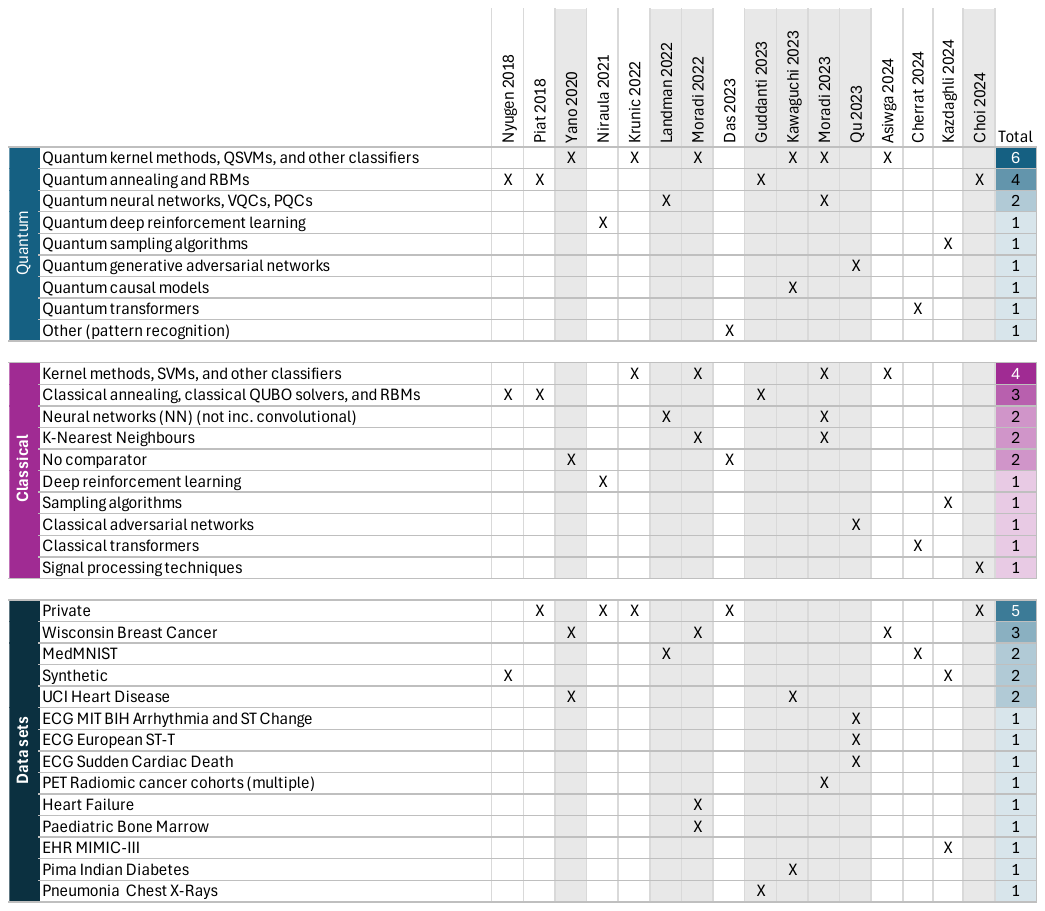}
\caption{ Overview of setting, interventions, and comparators for all \srsyn{} synthesized studies (columns), including studies with borderline or weak consensus (shaded gray columns). For each study, we indicate primary quantum interventions (top), classical comparators (middle), and digital health datasets (bottom). Quantum interventions deployed on non-digital health datasets are excluded. Interventions or comparators (rows) combine similar multiple-choice data extraction entries. QNNs exclude QCNNs and no VQCs / PQCs were found to contain mid-circuit measurements (MCM) or adaptive gates. A list of links to all sixteen studies, including any available codebases and datasets is compiled in \cref{app:databaselinks}.}\label{fig:tab:setting} 
\end{figure}

\begin{figure}[htbp]
\centering
\includegraphics[clip, trim=2cm 5.0cm 2cm 5.2cm, width=0.95\textwidth]{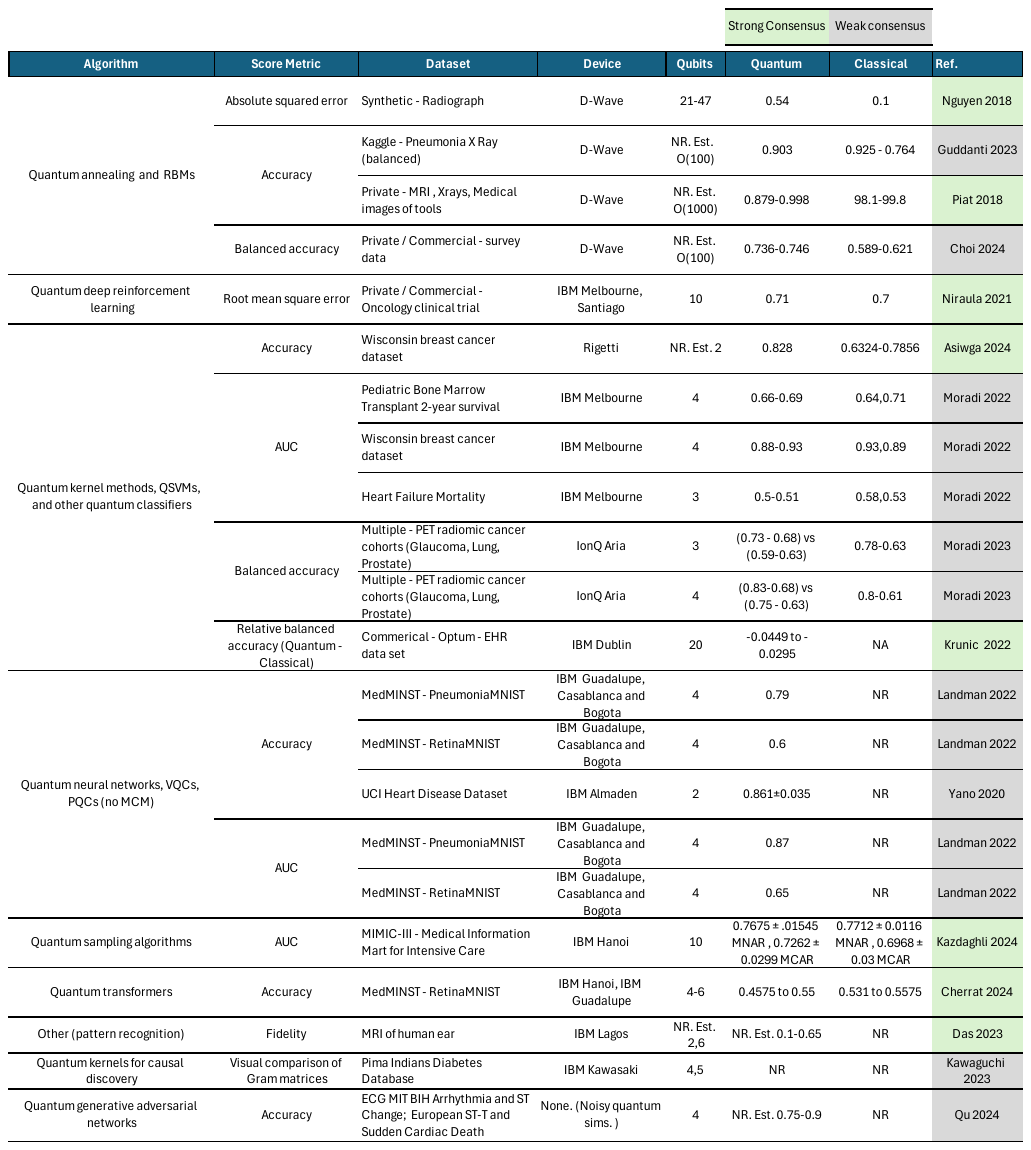}
\caption{Selected test scores (accuracy, AUC and squared error $\in [0,1]$) reported for quantum interventions deployed on quantum hardware using health data. Ranges reflected min and max test scores attained over different experimental configurations reported in each study; error bars for each test score are shown if originally reported. Synthesized studies with strong (weak) consensus are shaded in green (gray). Best possible scores for accuracy, fidelity and AUC metrics (unity) differ from squared error metrics (zero). Performance benchmarks that cannot be easily compared or interpreted alongside other metrics, or are not consistently reported, e.g. F1 scores and precision, are excluded for readability. NA: Not applicable. NR: Not reported. Est.: Estimates of numerical values implied from visual/graphical data.}\label{fig:tab:hscore} 
\end{figure}

The growth in the number of eligible articles on quantum algorithms for digital health seems almost exponential in \cref{fig:tab:metadata}(a). These applications are broadly categorized into diagnosis, predictive health, and `other' in \cref{fig:tab:metadata}(c). `Diagnosis' refers to an application that identifies, characterizes or labels current health data with the aim of supporting a clinical diagnosis e.g. classification of medical images or time varying signals. `Predictive' refers to an application that predicts future health information based on current health data, not necessarily to support formulation of a clinical diagnosis e.g. predicting drug efficacy or disease/risk factors. All remaining applications are group under `other` e.g. generating synthetic ECG signals based on EHR/EMR data. In \cref{fig:tab:metadata}(d),  we observe that the majority of studies did not enable code and data accessibility which are typically both required to enable tests of reproducibility.

\subsection{Empirical evidence from synthesized studies}

Nearly all of our synthesized studies were concerned with a learning task of performing a clinical diagnosis or a clinical prediction based on classical datasets. From a clinical perspective, all studies rationalized quantum algorithm design by citing other empirical literature. Any empirical rationale for the choice of quantum intervention did not necessarily refer back to comparable clinical settings: in most cases, it appeared that the matching between health datasets and quantum interventions was either ad-hoc, or one tried all possible quantum interventions in order to empirically discover the best performing models for a fixed health dataset. No QML applications were focussed on health service delivery, public health, and consumer health monitoring applications.  Only one study,  Qu (2023) \cite{Qu2023}, focussed on health-data analytics applications, namely, that QGANs may be beneficial in ameliorating issues of model collapse for synthetic data generation in digital health applications but these are untested at scale in both simulations and quantum hardware. No studies were related to improving efficacy or efficiency of health service delivery, e.g. optimization problems for patient flow, or operational cost-down in hospitals.

We find that quantum kernel methods and quantum annealing techniques dominate our synthesized evidence in \cref{fig:tab:setting} (top). The choice of quantum intervention typically then informs the choice of classical comparator (middle) within each study, and hence the distributions of quantum and classical algorithms are correlated. Finally we note that datasets (bottom) are not particularly clinically diverse and factorize into private and open-source datasets.  While EHR and hospital data are often private, the remaining datasets are all open-source. Most empirical evidence does not use electronic health records, but gravitates to a handful of open-source health databases. We thus find that the diversity of  applications investigated empirically is limited.

In \cref{fig:tab:hscore}, we report performance metrics from synthesized studies comparing quantum interventions with classical machine learning counterparts for different digital health applications. The choice of quantum algorithms again separates into two groups: annealing vs. gate-based techniques. Indeed, quantum annealing studies focus on digital health tasks that can be mapped to a QUBO problem, and are able to scale to problem sizes at least an order of magnitude larger than non-annealing quantum hardware in qubit number. On the other hand, gate-based non-annealing quantum hardware accommodates a broader range of QML algorithms, as shown by the remaining rows in \cref{fig:tab:hscore}, and a broader range of hardware platforms, such as trapped ions (IonQ) and superconducting qubits (Rigetti, IBM). However, hardware experiments in many instances are almost outdated e.g. IBM quantum processors (56.2\% of synthesized studies) are Falcon models or older despite the availability of processors with 100+ qubits since 2022. 

We compare quantum vs. classical machine learning by reporting selected metrics from sixteen synthesized studies in the remaining columns of \cref{fig:tab:hscore}. Of these, five studies do not provide sufficient information to address our review question. Das (2023) \cite{Das2023}, Kawaguchi (2023) \cite{Kawaguchi2023} and Qu (2023) \cite{Qu2023} do not report numerical metrics for quantum experiments, while Yano (2020) \cite{Yano2020} and  Landman (2022) \cite{Landman2022} report numerical quantum benchmarks but do not report a numerical classical comparator. The remaining 11 studies reported in \cref{fig:tab:hscore} reveal major issues in facilitating the comparison between quantum vs. classical interventions. There are three scientifically concerning flaws:

\begin{enumerate}
    \item {\bf No empirical evidence of performance scaling:} All quantum computing demonstrations, even in simulations, have not been carried out at scale. Leaving aside the issue of quantum advantage for classical datasets, empirical investigations on universal, gate-based quantum computers have not investigated performance as a function of increasing problem size or qubit number e.g. to $O(100)$ qubits. Even for these small-scale experiments on universal, gate-based platforms, only Krunic (2022) \cite{Krunic2022} plotted trend lines of performance vs. problem size / qubit number to establish empirical scaling behavior.  In all other cases, including annealing applications, algorithmic performance scaling was not established in ideal or noisy simulations or prior to running on quantum hardware.  \\

    \item {\bf Limited reporting of statistical uncertainties:} All studies provided limited or no discussion of how statistical fluctuations in test scores should be interpreted. Only Kazdaghli (2024) estimates and reports sample error bars for test score values \cite{Kazdaghli2024}, while Krunic (2022) proposed a technique to contextualize fluctuations in performance score to the underlying configuration of experiments using PTRI metrics \cite{Krunic2022} but in lieu of uncertainty analysis. In the absence of error bars, any differences in classical vs. quantum performance appeared to be statistically equivalent fluctuations for a range of configurations. \\

    \item {\bf Lack of noise characterization and impact of quantum hardware:} Most studies recognize the significantly large deterioration between ideal and actual quantum hardware performance due to the effect of noise. Despite this, studies compared quantum hardware performance mostly only with ideal simulations, rather than using noisy simulations or secondary data to provide insight into algorithm performance on hardware. Only two synthesized studies \cite{Das2023, Moradi2022} used noisy simulations to compare to hardware results in their analysis. When running algorithms on quantum hardware, only two studies \cite{Das2023, Moradi2023} explicitly  considered error mitigation. Of these, only one study used an application-agnostic error mitigation technique and distinguished between raw vs. mitigated results to contextualize the impact of noise \cite{Moradi2023}. All studies failed to take data to characterize performance of the underlying quantum hardware while running QML experiments. Consequently, these studies offer almost no insight into whether fluctuations in classical vs. quantum QML performance are entirely dominated by drift in performance of underlying quantum hardware.
\end{enumerate}

\begin{center}
    \fbox{ \parbox{\textwidth}{In all these studies, we conclude that the performance differentials between quantum and classical machine learning metrics for digital health reported in \cref{fig:tab:hscore} are negligible. Not only is empirical evidence difficult to synthesize and interpret, but the tabulated performance scores show no clear, consistent, statistically significant trend to support any empirical claims of quantum utility in digital health across a range of hardware platforms.}}
\end{center} 

\section{Discussion}\label{sec:discn}

We have discussed until this point why a meta-analysis of empirical evidence in synthesized studies is insufficient for claiming empirical quantum utility for quantum machine learning in digital health. This absence of empirical evidence may be understandable in a relatively new field where applications development may temporally lag new insights in quantum machine learning theory and new hardware capabilities. We now consider these observations on research methodology and themes below.

Even in a discipline that must rely on heuristics and empirical investigations, the majority of studies claim empirical quantum advantage but do not take into account realistic operating conditions in their analysis. The absence of noise characterization or noisy simulations to explain deviations of quantum hardware experiments from idealized conditions is particularly surprising. In 14 out of 16 studies, hardware results were compared to ideal simulations without any noise characterization or the use of noise simulations to contextualize results.  Of the two studies that used noise simulations, these simulations were limited to simple noise models. For example, in Qu (2023) \cite{Qu2023}, QGANs are used for synthetic heartbeat data generation. Ideal QGANs converged to accuracies ranging from 87.7\% -- 90.9\% for different types of heartbeat data. Standard noise simulations of bit-flip, phase-flip, amplitude damping and depolarizing noise at moderately strong levels reduced the range of accuracies to $\approx$ 75\% -- 90\%, where each noise model is individually considered. However, in realistic settings, these noise models are inadequate---at the very least, requiring a mixture of different error types. While similar noise simulations were used as evidence to show noise-robustness of QML methods, the limited nature of these noise models would not reflect realistic operating conditions.  

Indeed the field appears to lack empirical comparisons of quantum annealing vs. gate-based QML in regimes where quantum annealing is anticipated to have provable quantum advantage, e.g. for specific learning tasks such as binary classification. Four studies formulated learning tasks as QUBO problems  and used a quantum annealer. Two of these studies focussed on classification tasks using a support vector machine \cite{Piat2018, Guddanti2023}, which could be easily compared with a universal gate-based computer. The remaining two studies focussed on areas such as linear regression \cite{Choi2024} and data compression \cite{Nyugen2018} for which there does not appear to be provable computational advantage for quantum annealing. Since D-Wave architectures have been available for some time before newer quantum processors, two of these four studies represent our oldest publications dating back to 2018. All annealing studies are also subject to study considerations above, and our review did not find strong evidence of quantum annealers outperforming either newer gate-based, universal quantum computers or classical counterparts.

Only one synthesized study used electronic health records as opposed to generic digital health data. In Krunic (2022) \cite{Krunic2022}, electronic health records were used to perform kernel-based prediction of six-month persistence of rheumatoid arthritis patients on biologic therapies. Both quantum and classical kernels were compared for different configurations of number of features and number of samples of training data. The study  offers weak evidence of empirical quantum advantage when the configuration space is restricted to small dimensional datasets with a low number of features. Aside from directly using electronic health records, Kazdaghli (2024) \cite{Kazdaghli2024} focussed on using quantum interventions for data imputation in clinical data, applicable to the analysis of electronic health records, but also other types of clinical data, such as those used in clinical trials.  Meanwhile, some eligible but not synthesized studies discussed the use of quantum algorithms for securely pooling health data in federated learning applications \cite{Monnet2023May}. 

Some of the quantum algorithms encountered in this review cited significant improvements in data-encoding compared to typical approaches in \cref{sec:roleofdata}. Efficient image processing tasks are pursued using quantum transformers in Cherrat (2024) \cite{Cherrat2024} and Landman (2022) \cite{Landman2022}, while data imputation is pursued in Kazdaghli (2024) \cite{Kazdaghli2024}. While these applications in digital health differ, the underlying technologies in Landman (2022) \cite{Landman2022}, Cherrat (2024) \cite{Cherrat2024},  and Kazdaghli (2024) \cite{Kazdaghli2024} all rely on methods in Ref.~\cite{Kerenidis2022Jan} and appear to inherit favourable resource scaling from assuming specific hardware capabilities that do not exist generally. The underlying data encoders assume hardware can implement entangling gates on overlapping sets of qubits in parallel (as opposed to sequentially). This hardware capability is so-far only shown for small-scale trapped ions \cite{Grzesiak2020Jun} and it is not expected to scale to systems with large qubit number. 

Meanwhile health data consists of both continuous and discrete data and Yano (2020)~\cite{Yano2020} fills an existing gap in literature by looking at encoding of discrete variable data into quantum VQCs using Quantum Random Access Coding (QRAC). The authors argue that $ O(\log_2(d\tau))/2 $ improvement in circuit size complexity can be attained for discrete variable inputs, suggesting a two-fold improvement over amplitude scaling in \cref{fig:tab:encoding}. Nevertheless, the critical challenge for amplitude encoding strategies in QML is that linear runtime complexity can prohibit accessing super-polynomial advantage and this barrier is not addressed by the paper. 

Nearly all quantum algorithms were linear quantum models. Some theoretical evidence shows that linear quantum models will require exponentially more qubits than non-linear models \cite{Jerbi2023Jan}, and heuristic evidence shows that certain types of linear quantum models will not be useful for the analysis of classical datasets \cite{Bermejo2024Aug}. Even broadening to a larger pool of \srftr{} eligible studies, non-linear quantum models were not encountered. Of our synthesized studies, seven studies used linear quantum kernel methods including Moradi (2022, 2023) \cite{Moradi2022, Moradi2023}, Yano (2020) \cite{Yano2020}, Aswiga (2024) \cite{Aswiga2024},  Krunic (2022) \cite{Krunic2022} and  Kawaguchi (2023) \cite{Kawaguchi2023}. For non-kernel methods, the underlying technologies for Nirula (2021) \cite{Nirula2021}, Qu (2023) \cite{Qu2023} and Das (2023) \cite{Das2023} can be recast in linear form. Finally, the quantum transformers and data encoding strategies that yield favourable scaling properties in Cherrat (2024) \cite{Cherrat2024},  Landman (2022) \cite{Landman2022},  and Kazdaghli (2024) \cite{Kazdaghli2024} use methods developed in Ref.~\cite{Kerenidis2022Jan}. Aside from a variant proposed in Cherrat (2024), the data encoders and neural networks leveraged by these studies all appear to be described by the framework of linear quantum models. Indeed, the observed absence of clear, consistent performance trends in the empirical meta-analysis of the previous section could in part be explained by the underlying linear quantum models used for many of the studies. After publishing our pre-print, we were made aware of Ref. \cite{Monbroussou2024Sep} as an improvement of quantum methods in Landman 2022 and Cherrat 2024, consisting of a non-linear model in \cref{fig:qc}(d).  While Ref. \cite{Monbroussou2024Sep} fails our inclusion criteria, even its inclusion would not affect the overall conclusions of our review.

Despite the fact that all quantum models were trained by a supervised learning problem, no study explicitly characterized their optimization landscape. It is well known that optimization of supervised QML algorithms can be plagued by exponentially vanishing gradients (barren plateaus)\cite{Larocca2024May}, exponential concentration of kernel values \cite{Thanasilp2024Jun}, or exponentially concentrated local minima \cite{Anschuetz2022Dec}.  However only two out of sixteen synthesized studies mentioned optimization challenges associated with their proposed methods for supervised quantum machine learning. Here, Cherrat (2024) \cite{Cherrat2024}  and Landman (2022) \cite{Landman2022} stated that their proposed QML methods' structures may avoid barren plateaus. All studies failed to provide a systematic  characterization of their empirical optimization landscape, and the resources utilized by their chosen optimization protocol in practice. Meanwhile no substantial improvements are found in reducing shot number requirements for QML applications considered in this review. 

Finally, classical data preprocessing tasks are highly discretionary and impact on QML is poorly understood. There are two areas where data preprocessing is frequently used in QML: feature selection for kernel methods, and dimensionality reduction for data encoding. In feature selection, both the number of features \cite{Moradi2022}, and statistical significance of features were established using statistical tests \cite{Moradi2023} to aid kernel design. Meanwhile, dimensionality reduction is required to encode data on quantum hardware with limited qubit numbers, e.g. by cropping, PCA or LDA. However, the impact of dimensionality reduction on QML performance is unaddressed. For example, reducing images to $2^n$ length, where the number of qubits $n$ is small, risks creating duplication in training and testing datasets if two different full-sized images become identical after dimensionality reduction. Other preprocessing tasks include re-scaling, using statistical summaries, or transforming data, e.g. using Haralick features \cite{Haralick2007Nov} or Fourier methods, but there has been no characterization of the impact of these methods on investigations of empirical quantum advantage.  

By design, the medium of a systematic review cannot speculate on the general future potential for use-case discovery of QML in digital health. Many open questions remain in the use of quantum algorithms to solve classical inference problems. As examples: whether quantum kernels can offer an advantage in classifying real world data; the role of data vs. quantum model design for attaining quantum advantage, the effects of optimization protocols and noise on overall QML performance, or the existence of quantum sampling or optimization algorithms that could be of use in classical machine learning. However, we comment on research areas of immediate urgency and interest in the next section.

\section{Future outlook}\label{sec:fo}

Our review highlights that the language of quantum advantage, empirical quantum utility, speed-up, or resource efficiencies are poorly defined and frequently abused notions in literature. QML applications development could benefit from guidelines on what robust quantum vs. classical comparisons look like. Even leaving aside the issue of how to select the best classical comparator, a comprehensive review on the approaches for benchmarking quantum performance is thus of immediate urgency and interest.

Comparing computational cost improvements enabled by quantum algorithms can be a difficult task. As discussed in \cref{sec:bkgd},  computational costs are theoretically quantified by sample complexity number of queries) or time complexity (number of sequential operations). As examples, for sample complexity, one must ensure that information contained in each query or sample must be comparable across algorithms. Meanwhile for time complexity, operations contain  assumptions about hardware capabilities and these assumptions are not always explicitly stated nor consistent. As elucidated by our discussion on quantum transformers in our review, studies assume that groups of quantum operations can be parallelized. This assumption is not hardware-agnostic: certain quantum operations may be parallelized in some architectures and not in others. 

In the absence of theoretical assurances on complexity, we have seen in \cref{fig:tab:hscore} how empirical studies use performance metrics such as fidelity or accuracy to argue for the `utility' of quantum algorithms in information processing tasks.  We have discussed how arguments of empirical utility or advantage must demonstrate both scalability and robustness of performance. Since simulating 100 qubits can be within reach of classical computers, we argue that characterizing properties of QML algorithms as a function of system size is more important than reporting any single figure of merit at some arbitrary choice of system size. Secondly, relying solely on ideal simulations offer no insight into robustness, and one simple test of robustness is to understand and mitigate the impact of noise. Thirdly, we find that the choice of performance metrics in empirical studies is diverse, often ad-hoc, and limits how to perform meta-analysis of evidence in the field.  

\begin{figure}[htbp]
\centering
\includegraphics[clip, trim=1.2cm 21cm 1.0cm 1.5cm, width=1\textwidth]{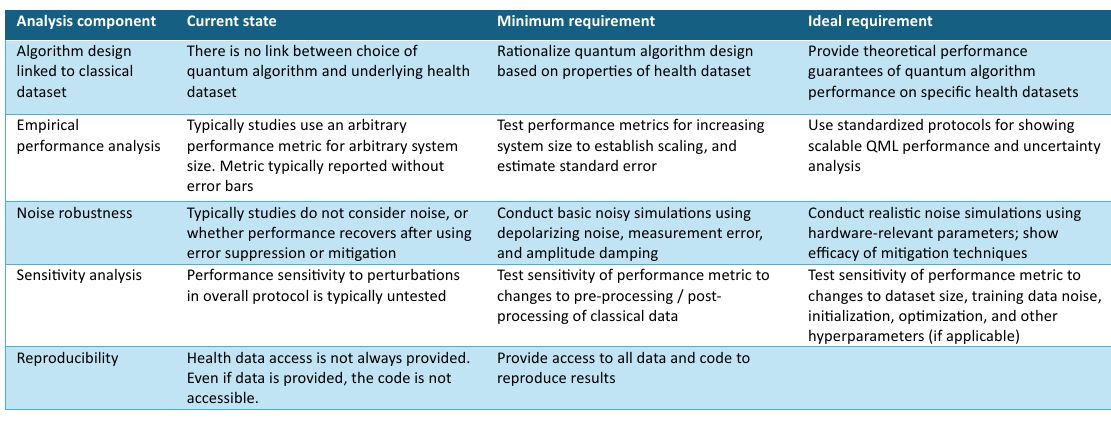}
\caption{Proposed minimal requirements for future QML study design, complementing the quality assessment framework used in our review (c.f. \cref{fig:tab:qascore}).}\label{fig:tab:futureoutlook} 
\end{figure} 

We summarize these and other considerations to specify minimum requirements for the robust analysis of quantum algorithms on classical datasets. Our proposed qualitative framework is presented in \cref{fig:tab:futureoutlook}. This framework is complementary to the quality analysis framework used in this review. In our framework, the minimum requirements outlined in \cref{fig:tab:futureoutlook} (column 3) can be immediately met. However, many of the ideal requirements in \cref{fig:tab:futureoutlook} (column 4) may require new research due to the nascency of the field. Indeed, the field has made some progress towards these ideal requirements: challenges of technical reproducibility in QML are discussed in Ref.~\cite{Bowles2024Mar} and quantum algorithm performance has been linked closely with hardware benchmarking in Ref.~\cite{Lubinski2021Oct}. However, research gaps continue to exist, for example, in the lack of a principled approach to link quantum algorithms to structures in classical data; to select appropriate performance metrics; to use noise characterization tools to set performance expectations and benchmark algorithm performance on hardware, and/or to compare quantum methods with a repository of best-in-class classical benchmarks  for industry subdomains. Further development and testing of the proposed framework is both urgent and important to urge better empirical evidence in our nascent field.  Finally, we note that our review methodology is  future-proof: by changing only the search period, we may provide a systematic update on the quality of research evidence for QML in digital health in future.

\section{Conclusion}\label{sec:con}

Digital health aims to transform access, affordability and quality of healthcare. As classical machine learning methods in health  approach commercialization, we find an exponentially growing number of studies advocating the use of QML in health. Our work is the first systematic review that examines the strength of empirical evidence to support these claims using a database of \srtabs{} studies. We find most applications are focussed on clinical decision support and comparatively little attention is given to health service delivery and public health use-cases. Of eligible studies, we appraise study quality yielding \srsyn{} robust studies which analyze QML applications in realistic operating environments. Despite this, we find that synthesized empirical evidence does not establish clear trends in performance benefits of QML algorithms over classical methods. Even leaving aside the issue of classical comparators, this synthesized evidence additionally does not establish scalability or robustness of QML performance. To this end, we propose minimal requirements for empirical studies for claiming empirical advantage for QML algorithms. We reiterate that enabling meaningful use-case discovery for QML in digital health requires new research to  rationalize the choice of QML structures on classical datasets, define appropriate benchmarks, and establish performance scaling under realistic operating conditions. An update to the search period of our review enables us to systematically track changes in this evidence base in the near future.
\bigskip 

\textbf{Author contributions:} R.S.G, T.E, S.S and J.D.P  were involved in study design, formulating research question, defining scope considerations, and inclusion/exclusion criteria. R.S.G, T.E, S.S developed search strategy, data extraction and quality appraisal templates. R.S.G, C.E.W, T.E and S.S participated in screening, data extraction, and resolution of conflicts. R.S.G performed data analysis. All authors contributed to writing the manuscript. 

\textbf{Acknowledgements:} S.S. and C.E.W were supported by the Australian Research Council Centre of Excellence for Engineered Quantum Systems (EQUS, CE170100009). R.S.G would like to thank K. Beer, R. Grewal, M. Riba, and R. Sweke for useful feedback.

\textbf{Competing interests:} All authors declare no financial or non-financial competing interests. 

\textbf{Data and code availability:}  The datasets generated by Covidence during data-extraction and underlying analysis code for the current study are freely available online via DOI: 10.6084/m9.figshare.27148386 \cite{Gupta2024QML}. A complete list of search strategies by database, inclusion/exclusion criteria, and data extraction are provided in Supplementary Materials.

\clearpage

\clearpage 
\appendix

\section{Quantum background}\label{sec:app:qnotation}

We briefly introduce notation and terminology associated with quantum algorithms.  Quantum states are represented by  the notation of a ket $|\cdot\rangle$. We will consider a specific type of quantum state, called a qubit (or quantum bit). At the extremal points, a qubit occupies the state associated classical binary bit $|0\rangle, |1\rangle$. However, a  qubit can also occupy values that are superposition states, 
\begin{align}
    |\psi\rangle := \alpha |0\rangle + \beta |1\rangle,
\end{align} where $\alpha, \beta$ are complex amplitudes satisfying $|\alpha|^2 + |\beta|^2 = 1$, and the linear combination of $|0\rangle, |1\rangle$  has a well-defined relative phase. Indeed, the relative phase relationships between quantum states in superposition allows quantum states to interfere, i.e. add or cancel out, in ways that may enhance computations.  When measured, quantum systems are also inherently probabilistic, i.e. in the above example, the measured qubit will be associated with a binary outcome $0$ ($1$) with probability $|\alpha|^2$ ($|\beta|^2$). For most applications, these quantum measurements are repeated for $N_{shots}$ number of times in order to build a statistical distribution over measurement outcomes from which useful quantities are inferred. 

We can demystify quantum computations by thinking of them as basic operations in linear algebra. Here, kets are represented as column vectors in a complex linear space, and quantum operations are represented by matrices, which act on states via matrix multiplication. For every quantum state ket, we associate a bra,  $\langle \psi|$, which can be thought of as a row vector containing the complex conjugate of elements of $|\psi\rangle$. To compare any two quantum states, $|\psi\rangle$ and $|\phi\rangle$, we compute their overlap as an inner product between $\langle \phi|$ and $|\psi\rangle$, denoted $\langle \phi |\psi\rangle$. To summarize quantum algorithms, it is useful to represent quantum states with matrices as well as vectors. The matrix representation of a quantum state $|\psi\rangle$ is the density matrix, $\rho:= |\psi\rangle \langle \psi|$, such that $\rho$ is an outer product. The density operator is required to satisfy three mathematical properties: unit trace ($\mathrm{Tr}[\rho] = 1$), hermicity ($ \rho=\rho^\dagger$) and positive semi-definiteness. These requirements ensure that measurement statistics can always be associated to quantum probability amplitudes (e.g. $\alpha, \beta$) of an underlying quantum state. In the above,  mathematical operations are denoted by the trace $\mathrm{Tr}$, $\langle \cdot \rangle$ denotes an inner product,  and $^\dagger$ denotes complex conjugation.

Nearly all transformations of a quantum state, including those within quantum machine learning algorithms in the presence of realistic noise, are called quantum channels. These channels, denoted $\eta$, can represent transformations such as quantum gates (rotations), quantum measurements, or the impact of noise on a quantum state interacting with its environment. For all these phenomena, we represent channels as, 
\begin{align}
    \eta (\rho) := \sum_i K_i \rho K_i^\dagger, \label{eqn:qchannel}
\end{align} where $K_i$ represents a quantum gate, measurement, or noise process. For the transformation $\eta (\rho)$ to output a physically valid quantum state, $\eta$ must satisfy mathematical properties (completely positivity and $0 \leq \mathrm{Tr}[\eta (\rho)] \leq 1$) or alternatively, $\sum_i K_i^\dagger K_i \leq 1$.

Quantum circuits only represent a simple subset of all possible quantum channels. By placing restrictions on what $\eta$ can represent, we can draw $\eta$ as a quantum circuit. A quantum circuit contains horizontal wires representing qubits where input states are shown as ket symbols to the left of each wire and temporal order progresses from left to right. Boxed symbols represent quantum gates or quantum measurements on relevant wires. The main restriction is that boxed quantum gates must be reversible rotations of quantum states, i.e. unitary operations satisfying $K^\dagger = K^{-1}$. With measurements pushed to the end, a quantum circuit typically represents a unitary quantum channel with only one term in the sum in \cref{eqn:qchannel}, i.e. $i\in \{1\}$, and therefore only an ensemble of quantum circuits can represent the collective action of noisy quantum circuits. The term `circuit size' represents the number of qubits, while `circuit depth' represents the number of time steps required to run the full circuit. Circuit depth assumes that quantum operations have been parallelized where possible and therefore refers to the minimum number of sequential steps, rather than a full tally of quantum gates in a circuit. While measurements are typically ignored until the very end, some measurements are performed `mid-circuit'. If these mid-circuit measurement outcomes, or measured qubits, are not used in subsequent processing, then they can be safely pushed to the end of the circuit. In other cases, a double-wire may be used to visually represent how outcomes of mid-circuit measurements can be used to change quantum operations `on the fly'.

Extracting information from a quantum computer is inherently statistical. In the case of qubits, all quantum measurements form a distribution of random `0' or `1' outcomes  i.e. binary outcomes of a Bernoulli trial. Typically there is some desired physical quantity, $\hat{O}$, whose average value must be computed from this distribution of quantum measurements. The quantum state $\rho$ is prepared, and the type (or basis) of quantum measurements is chosen according to $\hat{O}$. The statistical average (expectation value) of $\langle \hat{O}\rangle$ is then represented by,
\begin{align}
    f := \mathrm{Tr}[\rho \hat{O} ],
\end{align} where $f$ is estimated empirically by collecting experimental data on a quantum computer using $N_{shots}$, and the right-hand side of the equation is a mathematical description of the information extraction process. Since $\mathrm{Tr}[A^\dagger B] = \langle B, A \rangle $ represents an inner product, $f$ is some measure of the overlap between our quantum state $\rho$ and the desired operation $\hat{O}$, analogous to the overlap of quantum states represented as vectors. Indeed if $\rho = |\psi\rangle \langle \psi|$, then $\mathrm{Tr}[\rho \hat{O} ] = \mathrm{Tr}[|\psi \rangle \langle \psi | \hat{O} ] = \mathrm{Tr}[\langle \psi | \hat{O} |\psi \rangle] = \langle \psi | \hat{O} |\psi \rangle$ which is a scalar number representing the average value of $\hat{O}$ under $|\psi \rangle$. Computing this average requires repeatedly preparing $|\psi \rangle$, measuring this quantum circuit in the appropriate basis $N_{shots}$ number of times, and using classical post-processing of distribution of measurement outcomes. All quantum algorithms in this review extract information from quantum computers in this manner.

Let $f(x, \theta)$ be the average output of a quantum algorithm, and $\rho_0$ be the input quantum state, e.g. where all qubits are in their ground (zero) state, and where $(x, \theta)$ define classical inputs to a quantum algorithm. Here, $x$ represents one sample of real data with dimension $d$, $x \in \mathbb{R}^d$, for a dataset containing a total of $N$ data samples. Meanwhile, we also define tunable free parameters, $\theta$, that potentially could implement tunable quantum gates. The desired output information required from the algorithm is typically given by $\hat{O}$. We now discuss specific quantum algorithms that prepare quantum states, transform these states, and extract information from quantum computers in order to complete learning tasks.  

{\bf Quantum neural networks (QNNs)} consist of input data ($x$-dependent) and tunable ($\theta$-dependent) quantum operations. Generally, the output of a  QNN is,
\begin{align}
    f(x, \theta) &:=  \mathrm{Tr} \left[ U(x, \theta) \rho_0 U^\dagger(x, \theta)  \hat{O} \right] = \langle \rho_{x,\theta}, \hat{O} \rangle, 
\end{align} where the data ($x$-dependent) and tunable ($\theta$-dependent) components of QNNs cannot be separated. In the above, $U(x, \theta)$ represents a parameterized quantum gate which depends on data $x$ and tunable parameters $\theta$.  The equation above computes the overlap between information in the quantum state $ \rho_{x,\theta} = U(x, \theta) \rho_0 U^\dagger(x, \theta)$, and the desired output $\hat{O}$, using an inner product. 

In contrast, linear quantum models allow us to separate the $x$-dependent quantum operations and $\theta$-dependent quantum operations within the inner product \cite{Jerbi2023Jan}. In these models, we perform data encoding operations followed by tunable gates $V(\theta)$. As shown in \cref{fig:qc}(a), a linear QNN can be expressed by, 
\begin{align}
    f(x, \theta) &:=  \mathrm{Tr} \left[ U(x) \rho_0 U^\dagger(x) V^\dagger(\theta) \hat{O} V(\theta)\right] = \langle \rho_x, \hat{O}_\theta \rangle
\end{align} In the above, $\theta$ can take the form of any other classical parameters that are not $x$, data encoding is expressed by $\rho_x := U(x) \rho_0 U^\dagger(x)$, and the parameterized neural net is expressed as $\hat{O}_\theta:= V^\dagger(\theta) \hat{O} V(\theta)$.

With this structure, we can describe many quantum machine learning algorithms. For example, we can omit $\theta$ entirely, and recover sophisticated algorithms that focus on data encoding procedures. In kernel methods, $\theta$ is replaced by training data, and the algorithm output $f$ during prediction represents a linear combination of all training samples. Sometimes the action of $\rho$, $U(x)$ or $V(\theta)$ is non-trivially restricted to some subset of quantum states, yielding so-called quantum transformers. These choices are valid examples of linear quantum models, discussed below. 

{\bf Quantum kernel methods (QKMs)} are expressed as linear quantum models by replacing free tunable parameters $\theta$ by optimized linear combinations of training data given by ($\alpha, \mathcal{X}_T$). To see this, we redefine the second term in the inner product, $\hat{O}_\theta \equiv \hat{ O}_{\alpha, \mathcal{X}_T} :=  \sum_{t=1}^T \alpha_t \rho(x_t)$  for training data $x_t \in \mathcal{X}_T, t = 1, \hdots, T$. Substituting this expression into the inner product, the output of quantum kernel methods is \begin{align}
    f(x, \alpha, x_t): = \sum_{t=1}^T \alpha_t \langle \rho_x, \rho_{x_t} \rangle, 
\end{align} where the inner product compares the overlap between two quantum states parameterized by two data points, $x$ and training data sample, $x_t$, as  in \cref{fig:qc}(b). The weights, $\alpha$, are optimized during training. 

{\bf Quantum transformers}, such as those of synthesized studies in Cherrat (2024) and Landman (2022), use sophisticated data encoders and neural network structures.  Here, the data loaders ensure that the encoded state $\rho_x$ consists of all possible combinations of states where all but one qubit is nonzero, i.e. states with Hamming weight-1 like `00010' or `10000' but not `10100'. The action of the quantum neural network is then chosen to ensure that output superpositions of quantum states are also of Hamming weight of one. In particular, let $\rho_x$ denote the  Hamming weight-1 inputs, $\Lambda$ represents the restriction of linear algebra operations that preserve the weight of these states, and $V_\Lambda$ the quantum operation which implements $V_\Lambda |x\rangle  = |\Lambda x \rangle.$ Here, some choice of weight matrix $\Lambda$ enables one to compute linear  multiplication $\Lambda x$ using a quantum circuit. Choosing $\hat{O}_\Lambda \equiv V_\Lambda \hat{O} V_\Lambda  $ yields all the data loaders introduced in both Cherrat (2024) and Landman (2022). Similarly, one can add a trainable layer with parameters $\theta$ such that a quantum operation implements a trainable matrix multiplication, $V_{W}(\theta) |x\rangle  = | W(\theta) x \rangle$, where trainability of $W$ is made explicit in notation. For one of the algorithms, the quantum orthogonal transformer discussed in Cherrat (2024), we find that the output function for computing the so-called attention mechanism $A_{i,j}$ for two data patches $x_i, x_j$, is
\begin{align}
   A_{i,j} \equiv f(x_i, \theta,  x_j) &:= \mathrm{Tr}\left[ U^\dagger(x_j) V_{W}(\theta) U(x_i) \rho_0  U^\dagger(x_i) V^\dagger_{W}(\theta) U(x_j) \hat{O} \right] = \langle \rho_{x_i} , \hat{O}_{x_j,\theta} \rangle, \\
   \hat{O}_{x_j,\theta} &= V^\dagger_{W}(\theta) U(x_j) \hat{O} U^\dagger(x_j) V_{W}(\theta).
\end{align} In the above, the term $\hat{O}_{x_j,\theta}$ is reminiscent of a quantum kernel method since it depends on another data sample $x_j \neq x_i$, but also depends on a trainable matrix $W(\theta)$ that affects parameterization of  quantum gates.  Assuming two different patches, $x_i \neq x_j$, one can factorize the inner product by  grouping  $x_j, \theta$, and argue that the inner product remains linear in $x_i$. Unlike kernel methods, however, $x_j$ is not limited to the training dataset and consists of all pairwise combinations in the data. Indeed if $x_i \equiv x_j$, then the output function would be non-linear in $x_i$ in a manner similar to non-linear quantum models \cite{Jerbi2023Jan} such as quanutm data re-uploading classifiers (QDRCs) depicted for reference in \cref{fig:qc}(c)\cite{Perez-Salinas2020Feb}.

{\bf Quantum convolutional neural networks (QCNNs)} can similarly be understood as a tunable quantum channel $\tilde{\eta}_\theta $ that is composed of many smaller channels $\eta_{\theta_i}$, where quantum channels are introduced earlier. For each $i$-th tunable layer, the algorithm's structure can be written as $\tilde{\eta}_\theta (\rho) := \bigcirc_{i} \eta_{\theta_i}(\rho) = \hdots \eta_{\theta_3}(\eta_{\theta_2}(\eta_{\theta_1}(\rho))) \hdots$. In typical formulations such as that in  \cref{fig:qc}(d), in each $i$-th layer, we measure half of the remaining qubits in that layer, forcing these measured qubits to be reduced to classical bits. Consequently, these channels $\eta_{\theta_i}$ are defined on an increasingly smaller number of qubits as $i$ increases, until only one qubit is left. The channels for each layer $\eta_{\theta_i}$ are typically non-unitary, meaning that unlike quantum gates, these operations cannot be reversed or `undone'. An example of a non-unitary channel is where the measurement outcomes of a pooling layer dictate how gates are applied to the remaining qubits in the next layer \cite{Cong2019Dec}. If we want to extract $f(x, \theta) = \mathrm{Tr}\left[ \tilde{\eta}_\theta(\rho_x) \hat{O}\right]$, then in general it appears that tunable parameters and data-dependent operations cannot be separated for QCNNs. However, if mid-circuit QCNN measurements do not affect future quantum operations and can all be safely pushed to the end of the QCNN circuit, then averaging over mid-circuit measurements can be implemented entirely in classical post-processing and the layers $\eta_{\theta_i}$ can be unitary (i.e. $\eta_{\theta_i}(\rho):= V(\theta_i)\rho V^\dagger(\theta_i)$). In this regime, we recover a linear quantum model for QCNNs, 
\begin{align}
    f(x, \theta) = \mathrm{Tr}_{i \neq q}\left[ \rho_x \tilde{\eta}_{\theta}^{\dagger} (\hat{O})\right],
\end{align} where the trace over $i$  conveys that  all pooling layers are marginalized at the end except for the last remaining qubit, $q$. There is some heuristic evidence that linear QCNN models of this form are unlikely to be useful for the analysis of classical data \cite{Bermejo2024Aug}. 

{\bf Quantum causal modeling} is the use of quantum  algorithms to solve causal inference problems. In medical settings,  establishing causality between variables based on real-world medical data is an important classical learning task. If the direction of a certain causal relationship is known, then the causal effect, which represents the strength of the causal relationship, can be estimated via classical or quantum techniques. In this review, the quantum techniques used for causal inference were all linear quantum models of the form above. 

{\bf Quantum deep reinforcement learning networks} rely on the same underlying capabilities of a quantum neural network and the specific examples encountered in this review fit into the framework of linear quantum models. The only modification is the introduction of reinforcement learning, whereby a classical learning agent is trained to take optimal actions in a given environment to maximize a pre-determined reward function. Since the learning agent is classical, the underlying role of quantum technologies is the same as any other linear quantum model. 

{\bf Quantum generative adversarial networks} use the interaction between two artificial learning agents that have access to a quantum computer. The generator and discriminator are adversarial: the generator creates random, synthetic data samples with the goal of fooling the discriminator into believing it is real data, while the discriminator must assign a label of real or fake to each data sample.  The end point of this game is a generator that has been trained to create high quality synthetic samples such that the discriminator is forced to guess randomly, i.e. discriminator guesses correctly with 50\% probability. In the quantum versions of these algorithms, the data refers to quantum states of a quantum system, the generator has access to a quantum computer, and the discriminator can perform arbitrary quantum measurements. QGANs may demonstrate exponential quantum advantage for sufficiently high-dimensional quantum data \cite{Lloyd2018}. For classical datasets, no provable quantum advantage exists. The quantum circuits formed by QGANs in this review were also linear quantum models.

\section{List of synthesized studies, code and datasets \label{app:databaselinks}}

For sixteen synthesized studies, we report available open-source code and datasets, where available with NR./UR. indicating Not Reported / Upon Request.

{\small \begin{tabular}{ l c l l}
 Synthesized Study & Ref. & Code & Data \\ 
 \hline
 Nyugen 2018 &\cite{Nyugen2018} & NR./UR. & Synthetic \\  
 Piat 2018 &\cite{Piat2018} & NR./UR. & Private \\
 Yano 2020 &\cite{Yano2020} & NR./UR. & UCI MLR Heart Disease \href{https://archive.ics.uci.edu/ml/datasets/heart+Disease}{Data}.\\
 & & & UCI Wisconsin Breast Cancer \href{https://archive.ics.uci.edu/dataset/14/breast+cancer}{Data}.\\
 Niraula 2021 &\cite{Nirula2021} & NR./UR. & Private: RTOG0617 dataset collected under NCT00533949-D1 (c.f. \href{https://nctn-data-archive.nci.nih.gov/node/286}{NCTN}) \\
 Krunic 2022	&\cite{Krunic2022} & NR./UR. & Private: Optum de-identified EHR dataset \\
 Landman 2022 &\cite{Landman2022} & NR./UR. & MEDMINST \href{https://medmnist.com/}{Data}.\\
 Moradi 2022	&\cite{Moradi2022} & \href{https://github.com/sassan72/Quantum-Machine-learning}{Git repo}&  UCI Pediatric Bone Marrow transplant \href{https://archive.ics.uci.edu/dataset/565/bone+marrow+transplant+children}{Data}. \\ 
  & & & UCI Wisconsin Breast Cancer \href{https://archive.ics.uci.edu/dataset/14/breast+cancer}{Data}.\\
   & & & Kaggle Heart Failure \href{https://www.kaggle.com/andrewmvd/heart-failure-clinical-data}{Data}.\\
 Das 2023 &\cite{Das2023} & NR./UR. & Private: MRIs of human ear collected in laboratory\\
 Guddanti 2023 &\cite{Guddanti2023} & \href{https://github.com/Sai-sakunthala/Pneumonia-Detection-by-Binary-Classification}{Git repo}& Kaggle Pneumonia Chest X Ray \href{https://www.kaggle.com/datasets/pcbreviglieri/pneumonia-xray-images}{Data} \\
 Kawaguchi 2023 &\cite{Kawaguchi2023} & NR./UR. & Kaggle Pima Indian Diabetes \href{https://www.kaggle.com/datasets/uciml/pima-indians-diabetes-database}{Data}.\\
 & & & UCI MLR Heart Disease \href{https://archive.ics.uci.edu/ml/datasets/heart+Disease}{Data}.\\
 Moradi 2023 &\cite{Moradi2023} & \href{https://github.com/sassan72/learning-with-Quantum-machines?tab=readme-ov-file}{Git repo} & Papp L et. al. \href{http://link.springer.com/10.1007/s00259- 020-05140-y}{Data}. Grahovac M et. al. \href{https://link.springer.com/10.1007/s00259-023- 06127-1}{Data}. Papp L, Pötsch N et al. \href{https://pubmed.ncbi.nlm.nih.gov/29175980/}{Data}.\\
 Qu 2023 &\cite{Qu2023} & \href{https://github.com/VanSWK/QCGAN_ECG}{Git repo}&  \href{https://physionet.org/about/database/}{MIT-BIH Arrhythmia \& ST Change, European ST-T, Sudden Cardiac Death}.\\
 Asiwga 2024	&\cite{Aswiga2024} & NR./UR. & UCI Wisconsin Breast Cancer \href{https://archive.ics.uci.edu/dataset/14/breast+cancer}{Data}.\\
 Cherrat 2024 &\cite{Cherrat2024} & NR./UR. & MEDMINST \href{https://medmnist.com/}{Data}.\\
 Kazdaghli 2024 &\cite{Kazdaghli2024} & \href{https://github.com/AstraZeneca/dpp_imp/}{Git repo}& Synthetic, EHR MIMIC-III \href{https://www.nature.com/articles/sdata201635}{Data}.\\
 Choi 2024 &\cite{Choi2024} & \href{https://github.com/Jungguchoi/quantum_annealing_based_feature_selection}{Git repo}& Private survey datasets.  \\
 \end{tabular}}
 
\clearpage
\section{PRISMA Checklist} \label{sec:app:primsa}

\begin{figure}[htbp]
\centering
\includegraphics[clip, trim=2.24cm 2cm 2.24cm 1cm, width=0.9\textwidth]{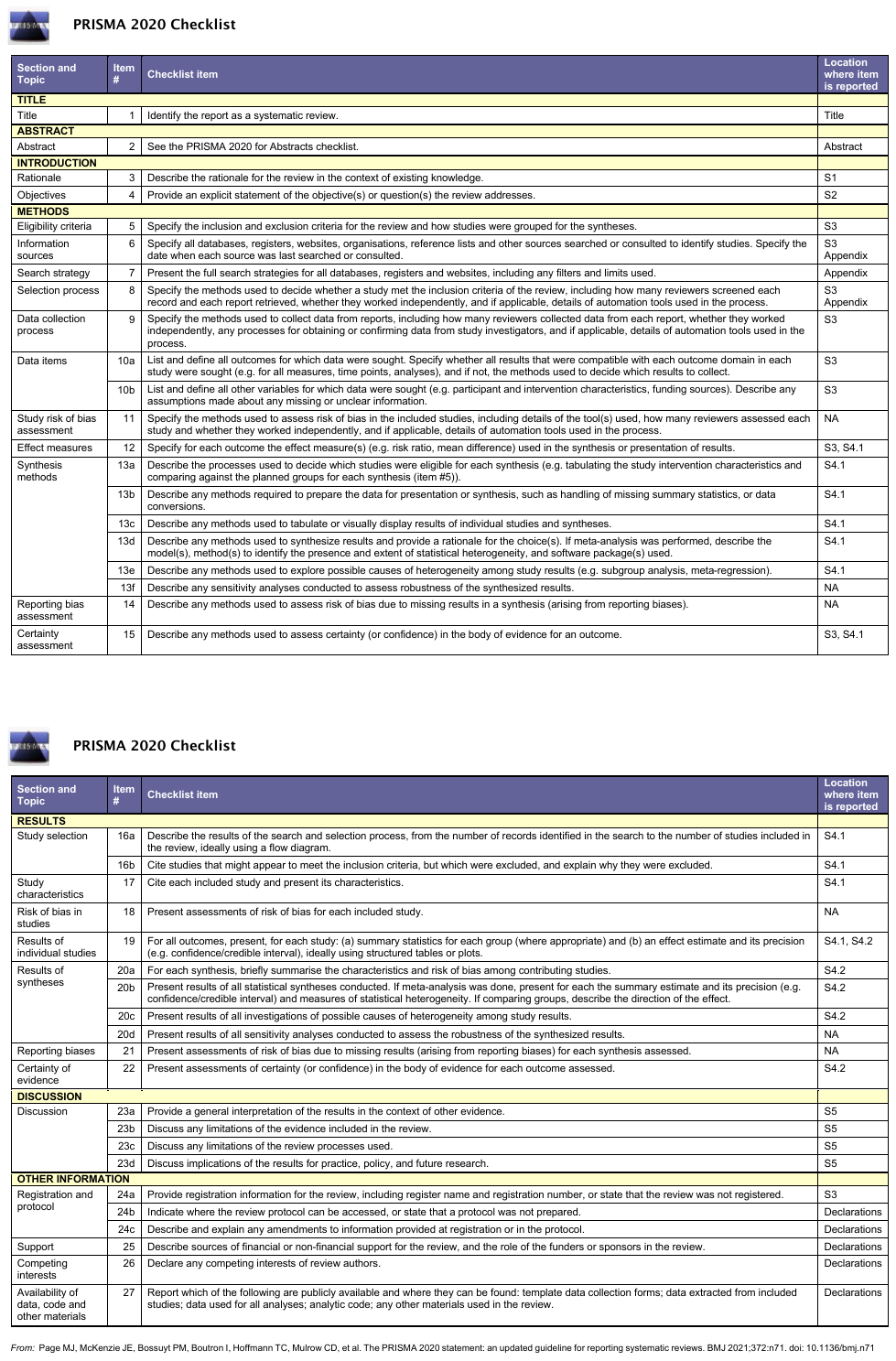}
\end{figure}

\begin{figure}[htbp]
\centering
\includegraphics[clip, trim=2cm 19cm 2cm 1cm, width=0.9\textwidth]{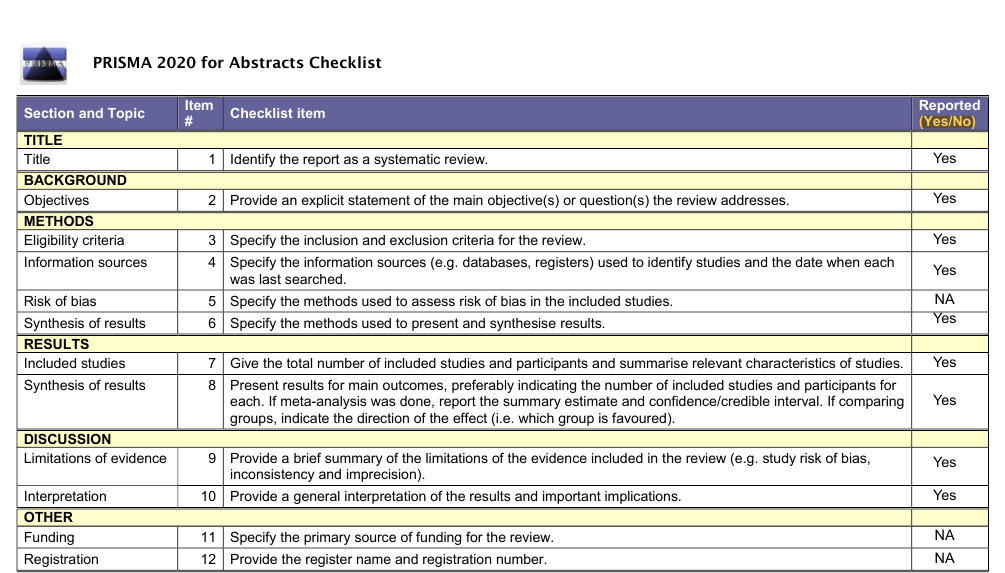}
\end{figure}

\FloatBarrier
\section{Nomenclature}\label{sec:app:nomenclature}

{\small \begin{figure}[htbp]
\centering
{\small \carlitoOsF \begin{tabular} {  m{2cm}  m{12cm} } 
\rowcolor{b0}\color{white} Abbrev. & \color{white} Definition \\
\hline
EHR & Electronic health records \\
\hline
EMR & Electronic medical records \\
\hline
GAN & Generative adversarial network \\
\hline
LDA & Linear discriminant analysis\\
\hline
PCA & Principal component analysis\\
\hline
PRISMA & Preferred reporting items for systematic reviews and meta-Analyses \\
\hline
PROSPERO & International prospective register of systematic reviews \\
\hline
PTRI & Phase space terrain ruggedness index, a proposed global metric for linking performance scores with a configuration space \cite{Krunic2022} \\
\hline
PQC &  Parameterized quantum circuit \\
\hline
QC &  Quantum computation\\
\hline
QCNN &  Quantum convolutional neural network algorithms\\
\hline
QEC &  Quantum error correction  \\
\hline
QEM &  Quantum error mitigation \\
\hline
QGAN &  Quantum generative adversarial methods\\
\hline
QKM &  Quantum kernel methods\\
\hline
QPIE &  Quantum probability image encoding \cite{Yao2018Jan}\\
\hline
QML &  Quantum machine learning algorithms\\
\hline
QNN &  Quantum neural network algorithms\\
\hline
QRAC & Quantum random access coding of bitstring length $m$ into $n$ qubits so that any 1 out of $m$ bits can be recovered with probability $p>1/2$\\
\hline
QRAM & Quantum random access memory  \\
\hline
QUBO & Quadratic unconstrained binary optimization \\
\hline
SVM &  Support vector machine \\
\hline
VQC &  Variational quantum circuit \\
\bottomrule
\end{tabular}}
\label{fig:tab:nomen}
\end{figure}}

\section{Search strategies}\label{sec:app:search}

Our database selection captures publishing practices in healthcare/medicine but also in quantum computing. While PubMed and Embase are popular in health, we additionally include IEEE and Scopus which typically index computer science, physics and engineering journals. Physics-centric publications, e.g. APS journals and Quantum, are indexed in Scopus. We additionally  search quantum physics preprint server arXiv (quant-ph) to capture recent and unindexed quantum computing and QML literature. 

\clearpage
\includepdf[pages=-]{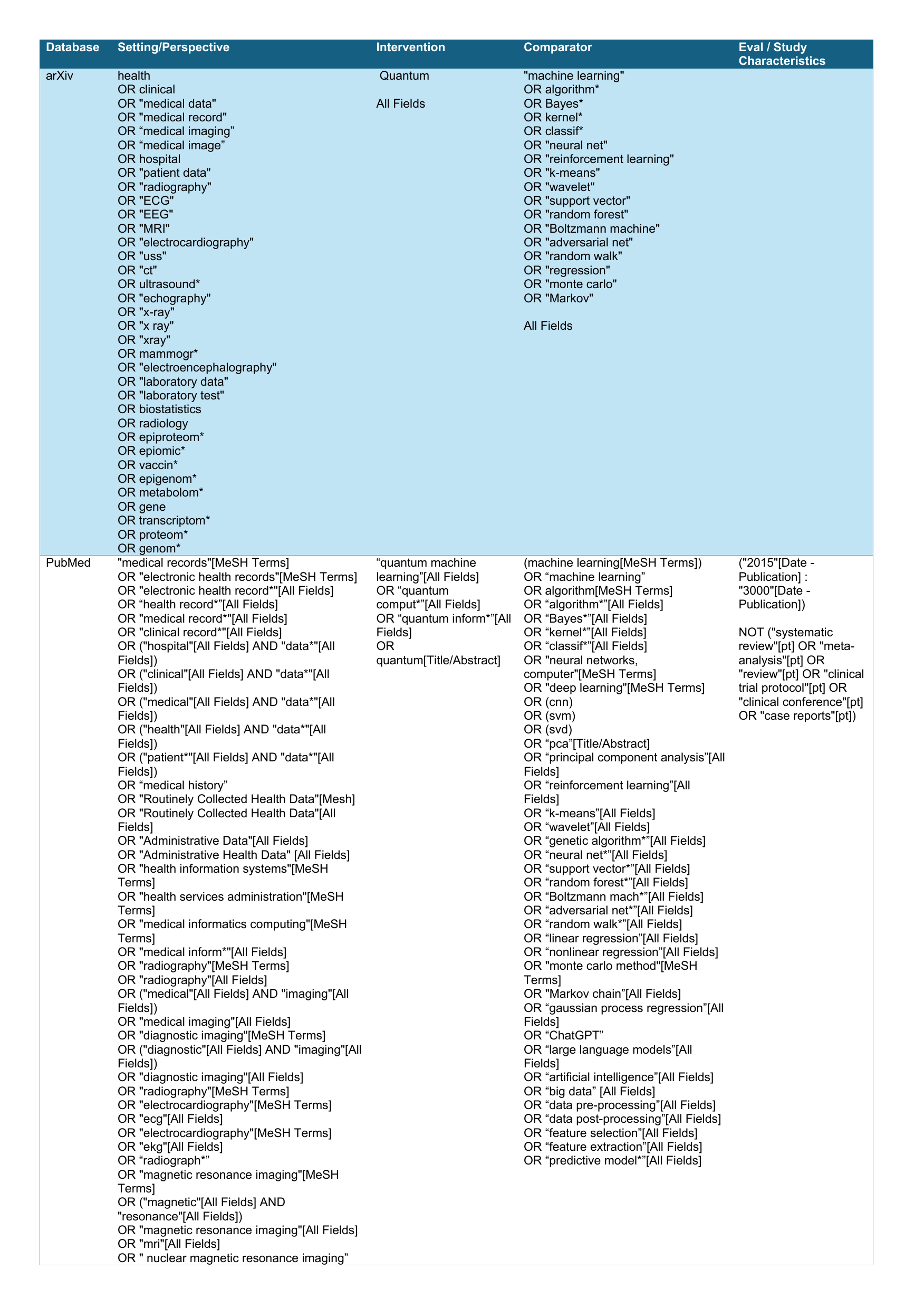}

\section{Data extraction template}\label{sec:app:dbextract}
\begin{figure}[htbp]
\centering
\includegraphics[clip, trim=1.3cm 1.5cm 1.3cm 1.5cm, width=0.922\textwidth]{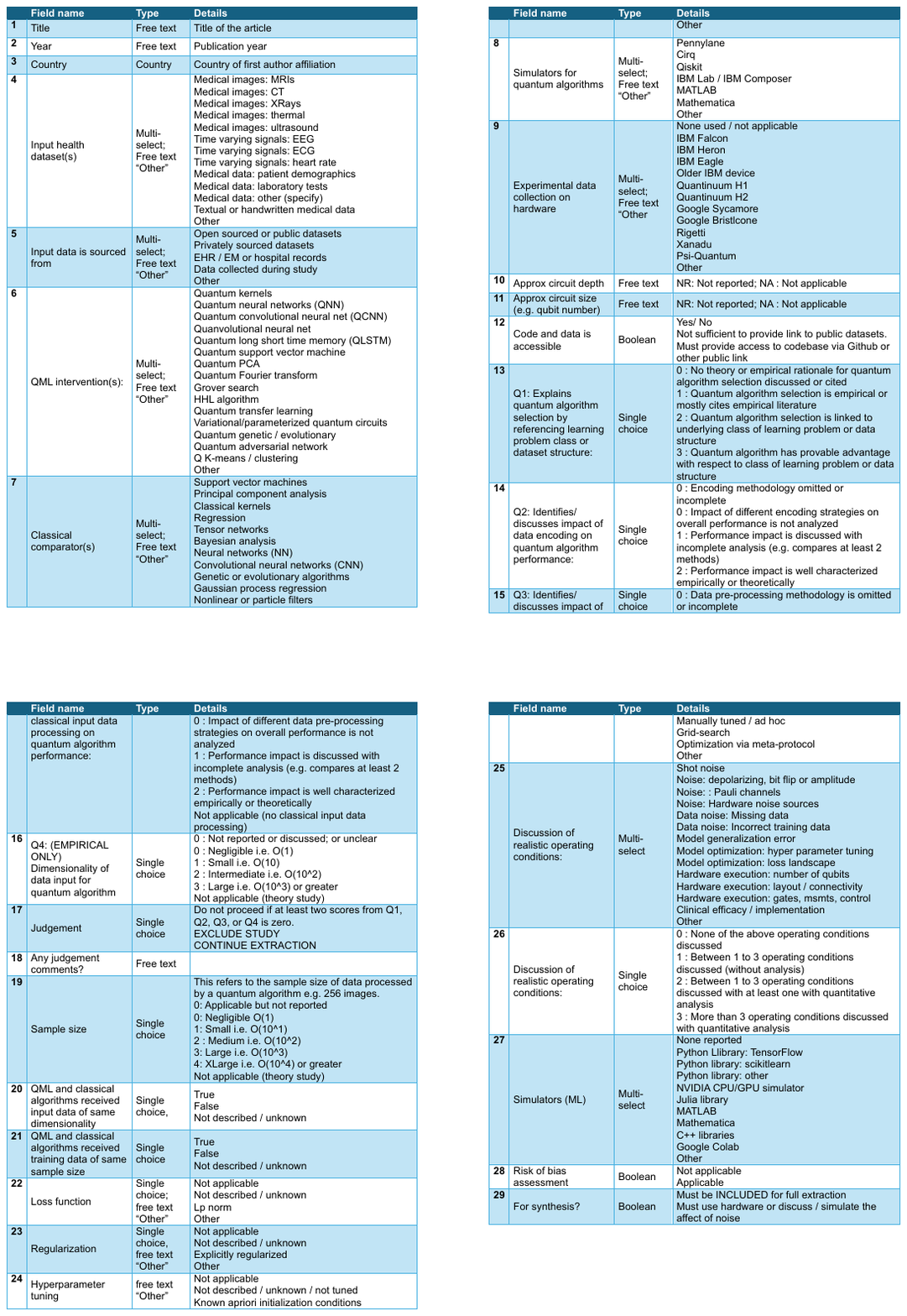}
\end{figure}


\end{document}